\documentclass[journal=jpcc,manuscript=article,layout=twocolumn]{achemso}
\setkeys{acs}{articletitle = true}
\usepackage[percent]{overpic}
\usepackage{color}
\usepackage[table]{xcolor}
\usepackage[version=3]{mhchem} 

\usepackage{bbold}
\usepackage{hyperref}


\author{\small Braden M. Weight}
\affiliation{\small Department of Physics and Astronomy, University of Rochester, Rochester, NY 14627, U.S.A.}
\alsoaffiliation{\small Theoretical Division, Los Alamos National Laboratory, Los Alamos, NM, 87545, U.S.A.}
\email{bweight@ur.rochester.edu}
\author{\small Sergei Tretiak}
\affiliation{\small Theoretical Division, Los Alamos National Laboratory, Los Alamos, NM, 87545, U.S.A.}
\author{\small Yu Zhang}
\affiliation{\small Theoretical Division, Los Alamos National Laboratory, Los Alamos, NM, 87545, U.S.A.}
\email{zhy@lanl.gov}

\title[An \textsf{achemso} demo]{\large A Diffusion Quantum Monte Carlo Approach to the Polaritonic Ground State}

\begin{document}


\begin{abstract}
{\small Making and using polaritonic states (i.e., hybrid electron-photon states) for chemical applications have recently become one of the most prominent and active fields that connects the communities of chemistry and quantum optics. Modeling of such polaritonic phenomena using\textit{ ab initio} approaches calls for new methodologies, leading to the reinvention of many commonly used electronic structure methods, such as Hartree-Fock, density functional, and coupled cluster theories. In this work, we explore the formally exact diffusion quantum Monte Carlo approach (DQMC) to obtain numerical solutions to the polaritonic ground state during the dissociation of the H$_2$ molecular system. We examine various electron-nuclear-photon properties throughout the dissociation, such as changes to the minimum of the cavity Born-Oppenheimer surface, the localization of the electronic wavefunction, and the average mode occupation. Finally, we  directly compare our results to that obtained with state-of-the-art, yet approximate, polaritonic coupled cluster approaches.
}
\end{abstract}
\\
\\
\\
\\
\\
\\
\\
\\
\\
\\
\\
\\



{\centering LA-UR-23-28443}\newline
{\small 
\section{Introduction}

Recent experimental studies of highly entangled light-matter states, known as polaritons, have been shown their ability to modify chemical reactions\cite{Nagarajan2021JACS,GarciaVidal2021S,Hutchison2012ACIE,Schwartz2011PRL,Ebbesen2016ACR,Sau2021ACIE,Thomas2019S,Thomas2016ACIE,Thomas2020N,Lather2019ACIE,hirai_molecularPOL_ChemRev2023,Simpkins2023CR} and physical properties\cite{Thomas2021,Vergauwe2019ACIE,Berghuis2020JPCC,Berghuis2022AP,xu_ultrafast_NatCommun2023,Deng2010RMP,rozenman_PolTransport_ACSPhot2018} of both low-\cite{luttgens_population_2021,mohl_trion-polariton_2018,graf_near-infrared_2016,graf_electrical_2017,son_energy_2022,Allen_SWCNTpol_JPCC2022} and high-dimensional\cite{} material systems, This has garnered a substantial interest in the theoretical community.\cite{mandal_QEDChemRev_Arxiv2022,Ruggenthaler2022Arxiv,Weight_PolaritonPCCP_2023Arxiv,Wang2021AP} Specifically, computational chemists have devoted the past few years toward the ``re-generation" of various many-body methods -- ubiquitously applied to pristine many-electron systems -- for use in the case of strongly correlated electron-photon systems.\cite{foley_abQEDREVIEW_Arxiv2023,Weight_PolaritonPCCP_2023Arxiv,mandal_QEDChemRev_Arxiv2022} All these approaches attempt to solve a non-relativistic quantum electrodynamical (QED) Hamiltonian for the coupled electron-nuclear-photon system for its eigenstates, usually within the cavity Born-Oppenheimer approximation,\cite{flick2017JCTC,schnappinger_scQEDHFManyMol_Arxiv2023,weight_abQED_JPCL2023} being the case for most conventional electronic structure methods. These modified self-consistent (sc) approaches include the scQED Hartree-Fock (scQED-HF),\cite{haugland2020PRX,schnappinger_scQEDHFManyMol_Arxiv2023} density functional theory (scQED-DFT),\cite{flick2018ACSPhotonics,pellegrini2015PRL,flick2018NanoPhotonics,Ruggenthaler2014PRA,Ruggenthaler2018NRC,Ruggenthaler2018NRC} and coupled cluster theory (scQED-CC) techniques\cite{haugland2020PRX,Haugland2021JCP,Pavosevic2022JACS,deprince2021JCP}, to name a few. Moreover, the analogous methods for excited state simulations have also been recently developed, such as time-dependent scQED-DFT (scQED-TDDFT)\cite{Flick2020JCP,yang2021JCP,yang_polGRADS_JCP2022,liebenthal2022JCP,liebenthal_mean-field_Arxiv2023,Vu_enhanced_JPCA2022} and equation of motion scQED-CC (scQED-EOMCCSD).\cite{mordovina2020PRR,deprince2021JCP,liebenthal2022JCP}

In many of these cases, the drawbacks of the original method are exacerbated in its scQED analogue. For example, to describe the electron-photon correlations in DFT approach, new exchange-correlation functionals needs to be constructed to explicitly account for such effects.\cite{flick2018ACSPhotonics,pellegrini2015PRL,Ruggenthaler2018NRC,Flick2018PRL,schafer2021PNAS,flick2022PRL} The early attempts at developing such functionals resulted in a dramatic reduction in the quality of treatment of bare electron-electron correlations.\cite{weight_abQED_JPCL2023,Pavosevic2022JACS,Haugland2021JCP,liebenthal_mean-field_Arxiv2023,Vu_enhanced_JPCA2022} While  efforts toward constructing improved functionals for the scQED-DFT approach are on-going with marked success,\cite{schafer2021PNAS,flick2022PRL} the ``exact'' effects of the cavity presence on the electronic subsystem, are only trustworthy up to the choice of electron-photon and electron-electron exchange-correlation functionals.

One promising approach is the scQED-CC method, which adds correlation on top of the scQED-HF approach. Here the electron-electron correlations are treated by including single and double excitations, which is known to provide  accurate results, even for highly correlated many-electron systems. In fact, the CCSD approach is exact for two-electron systems in the absence of the cavity since the single and double excitations comprise the full configuration interaction limit (up to the choice of basis set). This is no longer valid when the cavity is present, since the cavity photons can be excited to an arbitrary level due to their bosonic nature.\cite{mordovina2020PRR} Due to the large computational expense, the scQED-CC method has only been used for small molecular systems coupled to the cavity with typically one or two photonic excitations included. This approach is expected to provide reliable results, even for large light-matter coupling strengths.\cite{haugland2020PRX,Pavosevic2022JACS} However, such truncation of the photonic excitations has been shown to contradict the full configuration interaction limit in the strong coupling regimes for simple Hubbard model systems, even when using up to 10 photonic excitations.\cite{mordovina2020PRR} Notably, this high-level photonic treatment is usually numerically intractable for models describing realistic molecular systems.

In this work, we explore the polaritonic ground state of the H$_2$ molecular system coupled to a single quantized cavity mode. We solve the Pauli-Fierz QED Hamiltonian in the long-wavelength approximation using the diffusion quantum Monte Carlo (DQMC) approach. This methodology is formally exact for the two-electron H$_2$ molecular system, even when many quantized photonic modes are considered. We then directly compare with state-of-the-art polaritonic coupled cluster methods to highlight the inherent approximations in such schemes. Various properties are examined, such as the localization of the electronic wavefunction and changes to the polaritonic potential energy surface. We also explore a decomposition of the photonic wavefunctions into the number (or Fock) basis. This analysis highlights the discrepancy between the scQED-CCSD and DQMC results due to the inclusion of high-level photonic excitations present even in the ground state. Finally, we examine the average photon number as a function of the nuclear separation length. Our results suggest the DQMC scheme as another chemically relevant approach toward modeling molecular and material systems inside the cavity.

\section{ Methodology }
\subsection{Electronic Hamiltonian}
The molecular Hamiltonian,
\begin{equation}
    \hat{H}_\mathrm{M} = \hat{T}_\mathrm{N} + \hat{T}_\mathrm{el} + \hat{V} = \hat{T}_\mathrm{N} + \hat{H}_\mathrm{el},
\end{equation}
consists of the nuclear kinetic energy, $\hat{T}_\mathrm{N}$, the electronic kinetic energy, $\hat{T}_\mathrm{el}$, and all pairwise Coulomb interactions, $\hat{V} = \hat{V}_\mathrm{ee} + \hat{V}_\mathrm{eN} + \hat{V}_\mathrm{NN}$, between electrons $e$ and nuclei $N$. The electronic Hamiltonian, $\hat{H}_\mathrm{el} = \hat{H}_\mathrm{M} - \hat{T}_\mathrm{N}$, which is approximately solved by many electronic structure  software, can be written as an eigenvalue problem and yields the Born-Oppenheimer approximation for the electronic states parameterized by the nuclear positions,
\begin{equation}\label{EQ:HAM_EL}
    \hat{H}_\mathrm{el} |\psi_j \rangle = E_j |\psi_j \rangle.
\end{equation}
In this work, we consider a ``simple'' two-electron H$_2$ molecular system and its interatomic dissociation inside and outside a photonic cavity. In this case, the electronic Hamiltonian takes the form,
\begin{align}
    \hat{H}_\mathrm{el} &= \frac{\hat{\bf p}_1^2}{2} + \frac{\hat{\bf p}_2^2}{2} + \frac{1}{|\hat{\bf r}_1 - \hat{\bf r}_2|} - \frac{1}{|\hat{\bf r}_1 - {\bf R}_1|} - \frac{1}{|\hat{\bf r}_1 - {\bf R}_2|}\nonumber \\&- \frac{1}{|\hat{\bf r}_2 - {\bf R}_1|} - \frac{1}{|\hat{\bf r}_2 - {\bf R}_2|} + \frac{1}{|{\bf R}_1 - {\bf R}_2|},
\end{align}
where $\{{\bf R}_1, {\bf R}_2\}$ denote the nuclear coordinates, $\{\hat{\bf r}_1, \hat{\bf r}_2\}$ and $\{\hat{\bf p}_1, \hat{\bf p}_2\}$ are the position and momenta operators of the electrons, respectively. Solving Eq.~\ref{EQ:HAM_EL} yields the electronic adiabatic states $|\psi_j\rangle$, their respective  eigenenergies $E_j$, and all subsequent properties. Achieving exact solution has proven to be challenging, even for the state-of-the-art approximate many-body methods, such as density functional theory (DFT). Exact diagonalization of this Hamiltonian requires the solution of a six-dimensional Hilbert space and can be accomplished via coupled cluster singles and doubles (CCSD), which coincides with the full configuration interaction (FCI) limit for this two-electron system.

\subsection{Diffusion Monte Carlo}

\begin{figure}[t!]
\centering
\includegraphics[width=0.85\linewidth]{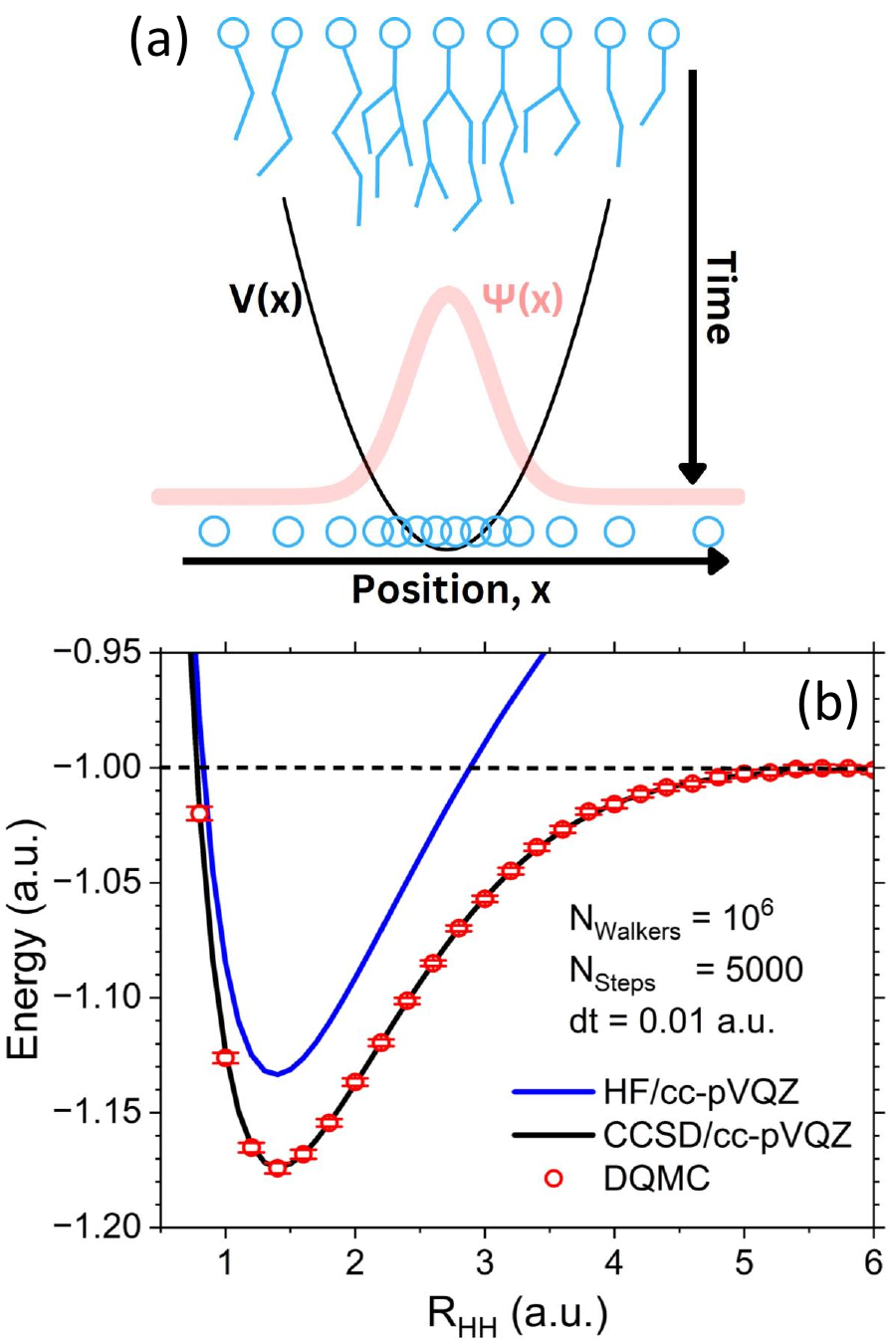}
  \caption{\footnotesize (a) A schematic of the diffusion Monte Carlo process. Gaussian random walkers are initially sampled from a uniform distribution (top blue circles) and move in time (top-to-bottom) by a Gaussian random walk. At each step, the random walker may be removed or duplicated according to the potential  energy landscape. (b) The ground state potential energy surface of the H$_2$ dissociation outside the cavity at various computational levels: Hartree-Fock (HF, blue curve), coupled cluster singles doubles (CCSD, black curve), and  diffusion quantum Monte Carlo (DQMC, red circles). Each simulation utilized initial 5000 timesteps to reach equilibrium followed by another 5000 timesteps with $d\tau$ = 0.01~a.u. Red error bars represent the standard deviation of the mean energy for each simulation at a fixed nuclear separation length. The horizontal dashed black line indicates the exact dissociation energy of $E_\mathrm{Diss.} = -1.0$ a.u.}
  \label{FIG:DQMC_SCHEME_NoCavComp}
\end{figure}

In many cases, the approximate solution to a high-dimensional integral problem can be achieved using a Monte Carlo approach, which leverages the concept of random variables in a high-dimensional space. Monte Carlo integration is ubiquitous in the community for both classical and quantum mechanical problems, when the number of degrees of freedom (DOFs) is large. One of the most common approaches is Markov-Chain Monte Carlo based on the Metropolis-Hastings algorithm,\cite{metropolis_monte_1949} which is related to the variational quantum Monte Carlo (VQMC),\cite{foulkes_quantum_2001} and is used to solve for the equilibrium distribution of the many-particle system.\cite{}

In this work, we focus on a particular quantum mechanical analogue called diffusion quantum Monte Carlo (DQMC).\cite{thijssen_computational_2007,martin_interacting_2016,johansson_qutip_2012,johansson_qutip_2013,micca_longo_unbiased_2021,umrigar_diffusion_1993,assaraf_diffusion_2000} In DQMC, the wavefunction is approximated by a basis of Gaussian random walkers whose 'motion' is defined by multiple applications of a short-time Green's function for the Schrodinger equation in imaginary time.

The imaginary time Schrodinger equation, in the six-dimensional real-space basis, can be written as,
\begin{equation}\label{EQ:SCH_EQ_IMAG_TIME}
    \frac{\partial}{\partial \tau} \psi({\bf r},\tau) = \bigg(\frac{1}{2}\boldsymbol{\nabla}_{\bf r}^2 + V({\bf r})\bigg)\psi({\bf r},\tau),
\end{equation}
where $\tau = it$ and $\frac{1}{2}\boldsymbol{\nabla}_{\bf r}^2 = \frac{1}{2}\boldsymbol{\nabla}_{{\bf r}_1}^2 + \frac{1}{2}\boldsymbol{\nabla}_{{\bf r}_2}^2$. Specifically, ${\bf r} = ({\bf r}_1, {\bf r}_2)$ is a single point in the six-dimensional configuration space of the two electrons in H$_2$. The formal solution of this equation can be written as a Green's function,
\begin{align}
    \psi({\bf r},\tau) = \int~d{\bf r}'~G({\bf r},{\bf r}', \tau) \psi({\bf r}',0),
\end{align}
This propagator $G({\bf r},{\bf r}', \tau)$ can be approximated by the Trotter-Suzuki splitting of the time-evolution operator as~\cite{Trotter1959AMS,SUZUKI1990PLA}
\begin{equation}
     e^{(\frac{1}{2}\boldsymbol{\nabla}_{\bf r}^2+V({\bf r}))d\tau}\approx e^{\frac{1}{2}\boldsymbol{\nabla}_{\bf r}^2d\tau} e^{V({\bf r})d\tau}
\end{equation} during a short-time interval $d\tau = \frac{\tau}{N_\mathrm{steps}}$ leading to the following Green's function approach,
\begin{align}
     G({\bf r},{\bf r}', \tau) =  \lim_{d\tau \to 0} \bigg[G_\mathrm{Diff.}({\bf r},{\bf r}', d\tau) G_\mathrm{Birth/Death}({\bf r},{\bf r}', d\tau)\bigg]^{N_\mathrm{steps}}.
\end{align}
The formal solutions to these Green's functions are
\begin{align}
    G_\mathrm{Diff.}({\bf r}, {\bf r}', d\tau) = e^{ -\frac{({\bf r} - {\bf r}')^2}{2d\tau} }\\
    G_\mathrm{Birth/Death}({\bf r}, {\bf r}', d\tau) = e^{-d\tau\frac{V({\bf r}) + V({\bf r}')}{2}}.
\end{align}
Here, and in the equations above, ${\bf r}$ and ${\bf r}'$ are two system configurations of the $3N_\mathrm{el}$ dimensional space (the nuclear DOFs are fixed), where $N_\mathrm{el}$ is the number of electrons ($N_\mathrm{el} = 2$ in this case). The first Green's function is the solution to the diffusion equation in free space (\textit{i.e.}, $\partial_\tau \psi = \frac{1}{2}\nabla^2 \psi$), which leads to an unbiased Gaussian random walk with a standard deviation $\sqrt{\tau}$.

The second propagator gives rise to an exponential probability of the Gaussian random walker itself, often called the Birth/Death algorithm. This propagator dictates the multiplication or destruction of a walker according to the probability distribution $\mathcal{P}_\mathrm{w} \sim G_\mathrm{Birth/Death}({\bf r}', {\bf r}, \tau)$, given the current $V({\bf r}')$ and previous $V({\bf r})$ total potential energies (for a single configuration of particles) of the system with configurations ${\bf r}'$ and ${\bf r}$, respectively. This term gives rise to a variable number of Gaussian random walkers, which can lead to an exponential increase (or decrease), of the number of walkers. 

For practical reasons, one introduces an energy shift, $E_\mathrm{T}$ in Eq.~\ref{EQ:HAM_EL} by replacing the energy eigenvalue $E_j$ with $E_j - E_\mathrm{T}$, where $E_\mathrm{T}$ is called the trial energy. The trial energy becomes an estimate of the exact ground state energy after sufficient simulation time, $ \lim_{\tau \to \infty} E_\mathrm{T}(\tau) \approx E_0$. As such, this parameter is dynamic and can be understood as a solution to a first-order rate equation for the number of Gaussian random walkers $N_\mathrm{w}$,
\begin{align}\label{EQ:TRIAL_ENERGY}
    &N_\mathrm{w}(\tau + d\tau) = N_\mathrm{w}(\tau) e^{-[E_\mathrm{T}(\tau) - E_\mathrm{T}(\tau+d\tau)]d\tau} = \bar{N}_\mathrm{w}\nonumber \\
    &\implies E_\mathrm{T}(\tau + d\tau) = E_\mathrm{T}(\tau) + \alpha \log\bigg(\frac{\bar{N}_\mathrm{w}}{N_\mathrm{w}(\tau)}\bigg)
\end{align}
where $\alpha$ and $\bar{N}_\mathrm{w}$ are parameters. $\bar{N}_\mathrm{w}$ is the target number of random walkers (taken to be 10$^6$ in this work). $N_\mathrm{w}(\tau = 0) = \bar{N}_\mathrm{w}$ to initiate the simulation. $\alpha$ is a parameter that controls the stiffness of the variation in the number of random walkers. In this work, we choose $\alpha = 0.01$ a.u., which damps the oscillations in the number of walkers while still allowing them to fluctuate according to the birth/death Green's function without encountering numerical issues such as zero or infinite walkers. A schematic representation of the DQMC method is provided in Fig.~\ref{FIG:DQMC_SCHEME_NoCavComp}a.

The DQMC scheme is formally exact for all nodeless ground states, which encompasses up to two Fermions (\textit{e.g.}, electrons) and, in principle, an infinite number of Bosons (\textit{e.g.}, photon modes). It is crucial to emphasize that all results presented in this work converge to the exact answer since we restrict our study to no more than two electrons. This convergence is ensured when a sufficiently small propagation timestep $d\tau$ is chosen and an adequate number of Gaussian random walkers $N_\mathrm{w}$ is used. Extensions of this scheme to ground states that have nodes (or phase changes) and to excited states have been well-studied for electronic systems. This augmentation often involves the fixed-node approximation, which necessitates \textit{a priori} knowledge of the wavefunction's nodal structure. This structure is typically derived from a Hartree-Fock Slater determinant or its post-Hartree-Fock counterparts.
 
\subsection{Pauli-Fierz Hamiltonian}
The coupling between light and molecular  DOFs can take many forms. In this work, we examine the interaction between the H$_2$ molecular system with a single quantized radiation mode (although there is no limit to the number of modes) using the Pauli-Fierz QED Hamiltonian within the Born-Oppenheimer approximation (\textit{i.e.}, neglecting the nuclear kinetic energy $\hat{T}_{\bf R}$). Mathematically this Pauli-Fierz Hamiltonian can be written as
  \begin{align}\label{EQ:H_PF}
    \hat{H}_\mathrm{PF} = &\hat{H}_\mathrm{el} + \omega_\mathrm{c}(\hat{a}^\dag\hat{a} + \frac{1}{2}) + \omega_\mathrm{c} A_0 (\hat{\boldsymbol{\mu}}\cdot \hat{e}) (\hat{a}^\dag + \hat{a})\nonumber \\&+ \omega_\mathrm{c} A_0^2 (\hat{\boldsymbol{\mu}}\cdot \hat{e})^2,
\end{align}
where the second term represents the bi-linear light-matter coupling term ($\hat{H}_\mathrm{el-ph}$), and the third term  denotes the dipole self-energy ($\hat{H}_\mathrm{DSE}$), DSE. The interactions between light and matter DOFs are controlled by the molecular dipole operator,
\begin{equation}
    \hat{\boldsymbol{\mu}}(\hat{{\bf r}}) = -\sum_p^{N_\mathrm{el}} \hat{{\bf r}}_p + \sum_I^{N_\mathrm{IONS}} Z_I {\bf R}_I,
\end{equation}
and its square,
\begin{equation}
    \hat{\boldsymbol{\mu}}^2(\hat{{\bf r}}) = \sum_{p,p'}^{N_\mathrm{el}} \hat{{\bf r}}_p \hat{{\bf r}}_{p'} - 2\sum_{p}^{N_\mathrm{el}}\sum_I^{N_\mathrm{IONS}} Z_I \hat{{\bf r}}_p {\bf R}_I  + \sum_{I,I'}^{N_\mathrm{IONS}} Z_I Z_{I'}{\bf R}_I {\bf R}_{I'},
\end{equation}
which encapsulates both one- and two-electron quadrupole-like terms. As will become clear in the next section, we require the light-matter Hamiltonian in the position representation for the cavity photon mode. Consequently, the Pauli-Fierz Hamiltonian can be rewritten as
\begin{align}\label{EQ:H_PF_qc}
    \hat{H}_\mathrm{PF} &= \hat{H}_\mathrm{el} + \frac{1}{2}\omega_\mathrm{c}^2 \hat{q}_\mathrm{c}^2 + \sqrt{2 \omega_\mathrm{c}^3} A_0 \hat{\mu} \hat{q}_\mathrm{c} + \omega_\mathrm{c} A_0^2 \hat{\mu}^2,
\end{align}
where $\hat{q}_\mathrm{c} = \sqrt{\frac{1}{2\omega_\mathrm{c}}}(\hat{a}^\dag + \hat{a})$ represents the position operator for the cavity photon.

\subsection{Polaritonic Diffusion Monte Carlo}

While DQMC has been proven to exactly solve the ground state of Bosonic systems, its extension toward entangled Boson-Fermion systems, particularly those arising from the strong coupling of molecular systems to light in optical or plasmonic cavities, remains under-explored. Nevertheless, extending this method to account for one or several quantized cavity or plasmonic modes, is relatively straightforward using a similar approach by decomposing the exact Green's function into multiple short-time propagators  for both the electronic and photonic DOFs .

Namely, we start from the imaginary-time  Schr\"odinger equation for the Pauli-Fierz Hamiltonian in the position representation for both the electrons ${\bf r}$ and cavity mode $q_\mathrm{c}$,
\begin{align}\label{EQ:SCH_EQ_IMAG_TIME_H_PL}
    &\frac{\partial}{\partial \tau} \psi({\bf r}, q_\mathrm{c}, \tau) \\&= \bigg(\frac{\nabla_{\bf r}^2}{2} + V({\bf r}) + \frac{\nabla_{ q_\mathrm{c}}^2}{2}  + V_\mathrm{el-ph}({\bf r},q_\mathrm{c}) + V_\mathrm{DSE}({\bf r})\bigg)\times\nonumber\\&~~~~~~~\psi({\bf r}, q_\mathrm{c}, \tau),\nonumber
\end{align}
where ${\bf r}$ again signifies all six real-space coordinates of two electrons. The many-body ``local'' dipole for a specific configuration ${\bf r}$, can be written as
\begin{equation}
    \boldsymbol{\mu}({\bf r}) = -\sum_p^{N_\mathrm{el}} {\bf r}_p + \sum_I^{N_\mathrm{IONS}} {\bf R}_I,
 \end{equation}
and  subsequent one- and two-electron quadrupole is
\begin{align}
    \boldsymbol{\mu}^2({\bf r}) = &\sum_{p, p'}^{N_\mathrm{el}} {\bf r}_p {\bf r}_{p'} - 2\sum_{p}^{N_\mathrm{el}}\sum_I^{N_\mathrm{IONS}} Z_I {\bf r}_p {\bf R}_I\nonumber\\&+ \sum_{I,I'}^{N_\mathrm{IONS}} Z_I Z_{I'}{\bf R}_I {\bf R}_{I'},
\end{align}
where $p$ and $p'$ label electrons and $I$ and $I'$ denote nuclei. It is worth noting that the nuclear charge $Z_I$ is 1, $N_\mathrm{IONS} = 2$, and $N_\mathrm{el} = 2$ for the H$_2$ cavity system.

 Utilizing the Trotter expansion, the kinetic energy of the electrons and photon can be split into one short-time Green's function propagator, while the potential terms can be  split into another set as
\begin{align}
    &e^{(\frac{\nabla_{\bf r}^2}{2} + V({\bf r}) + \frac{\nabla_{ q_\mathrm{c}}^2}{2}  + V_\mathrm{el-ph}({\bf r},q_\mathrm{c}) + V_\mathrm{DSE}({\bf r}))d\tau}\\
    &\approx e^{(\frac{\nabla_{\bf r}^2}{2}+\frac{\nabla_{q_\mathrm{c}}^2}{2})d\tau}e^{(V({\bf r}) + V_\mathrm{el-ph}({\bf r},q_\mathrm{c})+V_\mathrm{DSE}({\bf r}) - E_\mathrm{T})d\tau}\nonumber\\
    &\approx e^{(\frac{\nabla_{\bf r}^2}{2}+\frac{\nabla_{q_\mathrm{c}}^2}{2})d\tau}e^{(V_\mathrm{Total}({\bf r}, q_\mathrm{c}) - E_\mathrm{T})}
\end{align}
 over a short-time interval $d\tau$. Here,  $V_\mathrm{Total}({\bf r}, q_\mathrm{c}) = V({\bf r}) + V_\mathrm{el-ph}({\bf r},q_\mathrm{c}) + V_\mathrm{DSE}({\bf r})$ encompasses all the potential terms. We have also integrated the trial energy $E_\mathrm{T}$ directly into the above expression. This leads to the following Green's function approach for the  coupled electron-photon system,
 \begin{align}
     &G({\bf r}, q_\mathrm{c}, {\bf r}', q_\mathrm{c}', \tau) = \\& \lim_{d\tau \to 0} \bigg[G_\mathrm{Diff.}({\bf r}, q_\mathrm{c}, {\bf r}', q_\mathrm{c}', d\tau) G_\mathrm{Birth/Death}({\bf r}, q_\mathrm{c}, {\bf r}', q_\mathrm{c}', d\tau)\bigg]^{N_\mathrm{steps}}\nonumber.
\end{align}
In this case, the inclusion of the photonic DOF is formally analogous to an additional effective electronic one with a modified configurational potential energy $V({\bf r}) \rightarrow V({\bf r}) + V_\mathrm{el-ph}({\bf r}, q_\mathrm{c}) + V_\mathrm{DSE}({\bf r}) = V_\mathrm{Total}({\bf r}, q_\mathrm{c})$ dependent on its position $q_\mathrm{c}$ and the configurational electronic dipole $\mu({\bf r})$ as well as its square $\mu^2({\bf r})$.

 The formal solutions to these Green's functions are 
\begin{align}
    &G_\mathrm{Diff.}({\bf r}, q_\mathrm{c}, {\bf r}', q_\mathrm{c}', d\tau) = e^{ -\frac{|{\bf r} - {\bf r}'|^2}{2d\tau} } e^{-\frac{(q_\mathrm{c} - q_\mathrm{c}')^2}{2d\tau}}\\
    &G_\mathrm{Birth/Death}({\bf r}, q_\mathrm{c}, {\bf r}', q_\mathrm{c}', d\tau) =\nonumber\\&~~~~~e^{-d\tau\big(\frac{V_\mathrm{Total}({\bf r}, q_\mathrm{c}) - V_\mathrm{Total}({\bf r}', q_\mathrm{c}')}{2} - E_\mathrm{T}\big)}.
\end{align}
 Updating the trial energy $E_\mathrm{T}$ follows the same procedure as with photon-free propagation (Eq.~\ref{EQ:TRIAL_ENERGY}). The coupled electron-photon wavefunction is constructed by binning Gaussian random walkers at each timestep into a set of equally sized histograms (\textit{i.e.}, shared by all timesteps) such that the creation or destruction of walkers does not affect the histogram binning. Normalization is enforced at the end of the simulation.

\subsection{Computational Details}

 For clarity, we present a step-by-step breakdown of the proposed algorithm:

 \begin{enumerate}
      
    \item Generate initiate configurations by uniformly sampling all DOFs (Electronic Array Size:   $N_\mathrm{w}$ walkers $\times$ $N_\mathrm{el}$ electrons $\times$ 3  dimensions; Photonic Array Size: $N_\mathrm{w}$ walkers $\times$ $N_\mathrm{modes}$ ) over a wide enough range, taking special case to sample initial configurations well beyond the nuclear distribution.

     \item  Evaluate the total potential energy of the system  for each configuration $V_\mathrm{Total}({\bf r}, q_\mathrm{c})$. 

     \item Displace all walkers by sampling a Gaussian distribution with a standard deviation of $\sqrt{d\tau}$.

     \item  Evaluate the total potential energy $V_\mathrm{Total}({\bf r}, q_\mathrm{c})$ for the updated configurations.

     \item Update the trial energy $E_\mathrm{T}$ according to Eq.~\ref{EQ:TRIAL_ENERGY}.

     \item Calculate walker probabilities as:
     {\scriptsize
     \begin{equation}
        \mathcal{P}_\mathrm{w} = \mathrm{exp}\left[-d\tau \left( \frac{V_\mathrm{Total}({\bf r}, q_\mathrm{c}) + V_\mathrm{Total}({\bf r}', q_\mathrm{c'})}{2} - E_\mathrm{T} \right)\right].
    \end{equation}
    }

     \item For each walker, compare its probability $\mathcal{P}_\mathrm{w}$ to a unique uniform random number $\xi$ and perform one of the following actions
    \begin{enumerate}
        \item Kill the walker if $\mathcal{P}_\mathrm{w} < \xi$.
        \item Retain the walker if $\xi < \mathcal{P}_\mathrm{w} < 1$.
        \item Clone/duplicate the walker if $\mathcal{P}_\mathrm{w} > 1$.
    \end{enumerate}

     \item  Repeat steps 3-5 until the requested number of steps is completed.
     
    \item Repeat steps 2-6 using the final configurations as the initial configurations for the primary or ``production'' simulation.
     Save all average energies and  accrue configurational histograms (\textit{i.e.}, the wavefunction) during the production simulation.
\end{enumerate}

Unless otherwise noted, all data reported here use $N_\mathrm{w} = 10^6$ walkers, $N_\mathrm{steps}$ = 5000 steps for the equilibration as well as for the production simulations, $N_\mathrm{el} = 2$ electrons, and $N_\mathrm{modes} = 1$ cavity mode. We adopt a numerical timestep of $d\tau = 0.01$ a.u. and a population parameter of $\alpha = 0.01$ a.u. Our implementation is publicly available on GitHub: \href{https://github.com/bradenmweight/QuantumMonteCarlo}{https://github.com/bradenmweight/ QuantumMonteCarlo}.
\section{Results and Discussion}
\subsection{Bare H$_2$ Dissociation}

The dissociation curve of bare H$_2$ has been a common benchmark for new many-body methods as well as new density functionals for its simultaneous simplicity and complexity. Given the two-electron constitution, the Born-Oppenheimer surface can be computed exactly using  CCSD in the complete basis limit. In contrast, approximate methods, such as Hartree-Fock confined to the mean-field level of electron-electron correlation, perform very poorly for the H$_2$ dissociation. This can be improved using the broken-symmetry solution (\textit{i.e.}, performing an unrestricted self-consistent field  Hartree-Fock analog). However, such a solution, while improving the energy landscape of the mean-field solution, fails to capture precise symmetries of the electron orbitals. Hence, the physical nature of the wavefunction still requires further corrections.

From this perspective, CCSD and HF results comprise the best and worst limits of \textit{ab initio} approaches toward the H$_2$ dissociation, respectively, and will serve as reference for the current diffusion quantum Monte Carlo (DQMC) study. Figure~\ref{FIG:DQMC_SCHEME_NoCavComp}b  presents the results of the HF, CCSD and DQMC methods for the pristine (\textit{i.e.}, no cavity) H$_2$ dissociation. The HF and CCSD  results are obtained using the PySCF\cite{sun_pyscf_WIRES2018,sun_PySCF_JCP2020} electronic structure package using the cc-pVQZ basis set. For the DQMC results, the standard deviations of the mean energy are shown as vertical error bars for each nuclear separation length.

In this case, we find that the DQMC approach  agrees well with the reference CCSD results.  In the case of two electrons, CCSD is expected to perform well, contingent solely on the choice of basis set. However, introduction of the cavity photon mode when incorporating polaritons, leads to harsh approximations in the treatment of both bare photonic excitation and coupled excitations.  These truncations in the polaritonic CCSD (or scQED-CCSD) have been comprehensively discussed in existing literature.\cite{mordovina2020PRR,haugland2020PRX,Haugland2021JCP,deprince2021JCP,liebenthal2022JCP,Pavosevic2022JACS,mandal_QEDChemRev_Arxiv2022,Weight_PolaritonPCCP_2023Arxiv}.

\subsection{Polaritonic Dissociation Curves}

\begin{figure}[t!]
\centering
\includegraphics[width=1.1\linewidth]{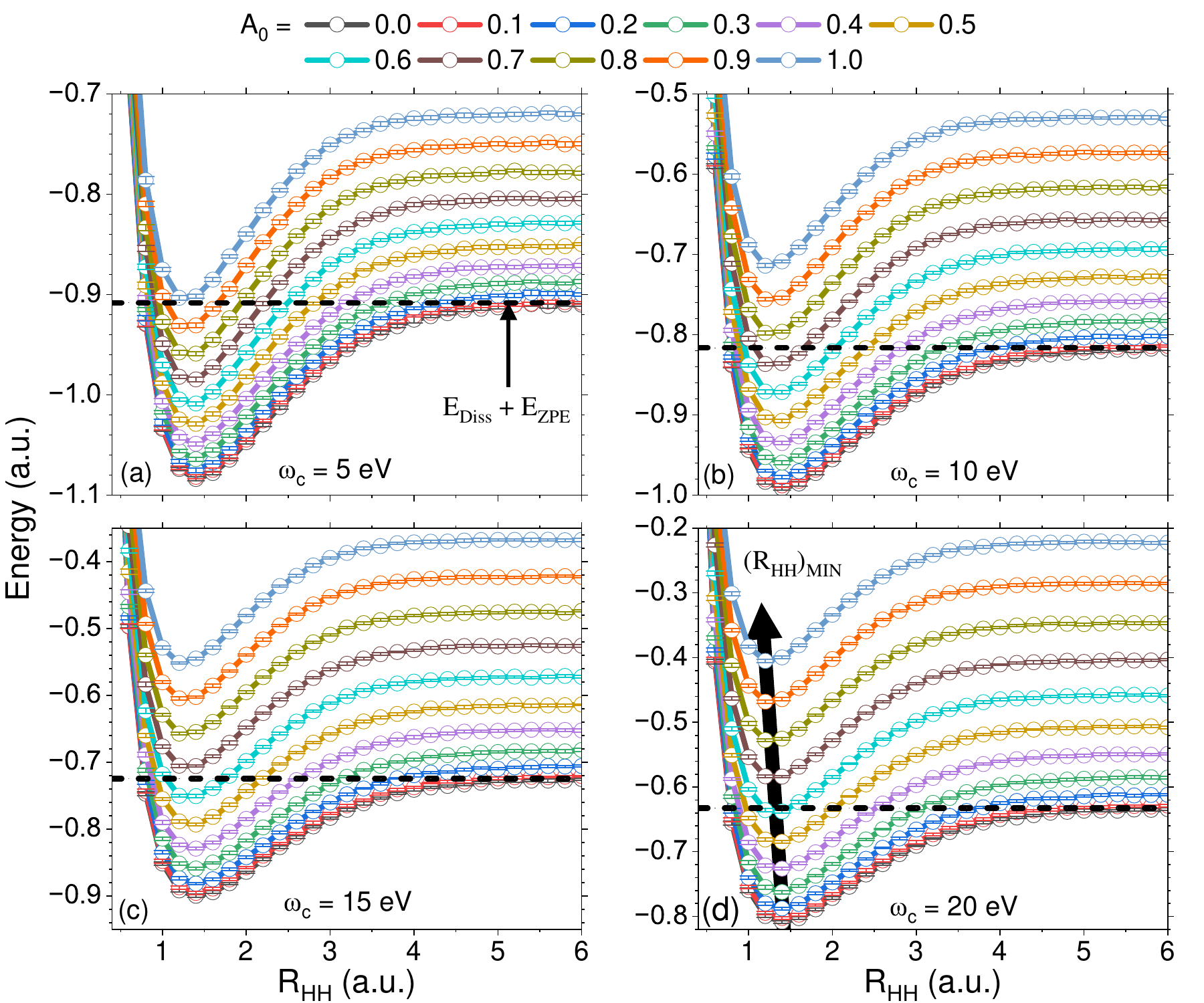}
  \caption{\footnotesize  Potential energy surfaces for the ground state H$_2$  dissociation at various coupling strengths $A_0$ = 0.0, 0.1, 0.2, 0.3, 0.4, and 0.5 a.u. for four cavity frequencies $\omega_\mathrm{c}$ = 5.0 (a), 10.0 (b), 15.0 (c), and 20.0 (d) eV. Vertical error bars represent the standard deviation of the mean energy. The horizontal dashed line references the expected dissociation energy with vanishing light-matter coupling calculated as  $E_\mathrm{Diss} + \frac{1}{2}\omega_\mathrm{c}$. The cavity polarization is parallel to the bond axis. The black arrow in panel (d) highlights the shift of the minimum energy point $(R_\mathrm{HH})_\mathrm{MIN}$  in the potential energy surfaces to lower values with increasing light-matter coupling strength $A_0$.
  }
  \label{FIG:PES}
\end{figure}

\begin{figure}[t!]
\centering
\includegraphics[width=0.7\linewidth]{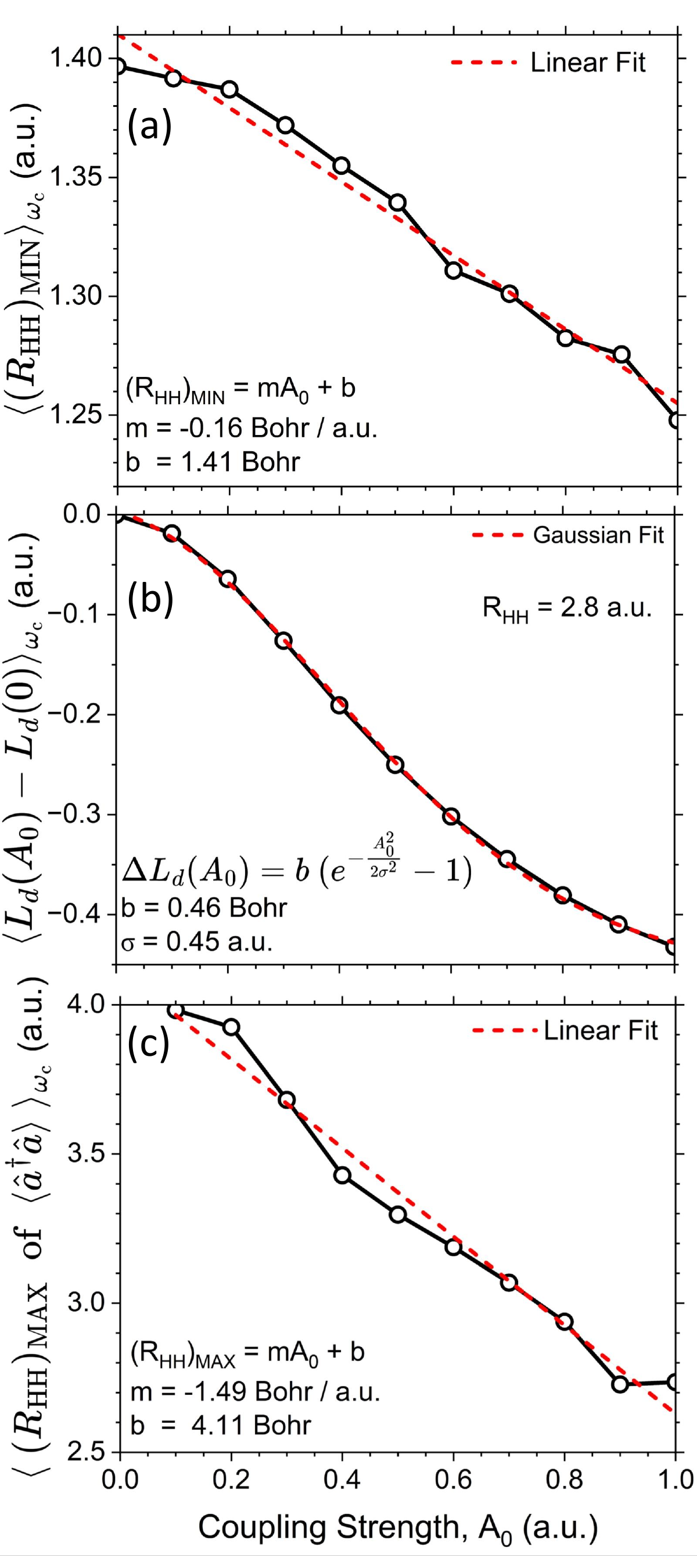}
  \caption{\footnotesize (a) The location of the ground state potential energy surface minimum, (b) the difference in wavefunction localization, and (c) the location of the maximum photon number as function of the light-matter coupling strength $A_0$. $\langle \cdots \rangle_{\omega_\mathrm{c}}$ indicates an average over cavity frequencies $\omega_\mathrm{c}$ = 5.0, 10.0, 15.0, and 20.0 eV. Panels (a) and (c) are accompanied with a linear fit (red line), and panel (b) is shown with a Gaussian fit (red curve).
  }
  \label{FIG:PROPERTIES_3PANEL}
\end{figure}

Figure~\ref{FIG:PES}  presents the primary results of this work: the polaritonic potential energy surfaces of the  H$_2$ dissociation at varied light-matter coupling  strengths $A_0$ (colors) and cavity frequencies $\omega_\mathrm{c}$. For each simulation (\textit{i.e.}, symbol), the standard deviation of the mean energy is shown by a vertical error bar. The horizontal dashed line indicates the expected dissociation energy at zero light-matter coupling , which is computed as the dissociation energy outside the cavity $E_\mathrm{Diss}$ and the zero-point energy of the cavity mode $E_\mathrm{ZPE} = \frac{1}{2}\omega_\mathrm{c}$.

 Two prominent features can be observed immediately: (I)  the vertical energy shift with increasing light-matter coupling strength $A_0$ and (II) the shift of the minima to lower values of $R_\mathrm{HH}$. This first observation is expected, since the primary contribution to the ground state energy is given by the DSE term, which only adds positive values to the total potential energy for each configuration. The second observation is, however, more interesting, since even small changes to such adiabatic surfaces may give rise to a wide range of modified chemistry. For example, the bond stiffness during a chemical reaction, the non-adiabatic couplings between electronic states in photoexcited processes. These phenomena can be monitored by specific spectroscopic signatures, either in the ground (\textit{e.g.}, infrared/Raman) or excited states (\textit{e.g.}, absorption/emission).  

The shift of the minimum in the potential energy surface as a function of the light-matter coupling strength $A_0$ is presented in Fig. ~\ref{FIG:PROPERTIES_3PANEL}a. The results are plotted as an average over all four cavity frequencies ($\omega_\mathrm{c}$ = 5.0, 10.0, 15.0, and 20.0 eV), given the minimal frequency dependency observed. A linear fit  to the data shows a slope of -0.16 Bohr / a.u. over this range of light-matter coupling $A_0$, leading to an overall reduction in nuclear separation of roughly 0.15 Bohr. Even this relatively large amount would result in substantial changes to a local chemistry. 

\begin{figure}[t!]
\centering
\includegraphics[width=1.0\linewidth]{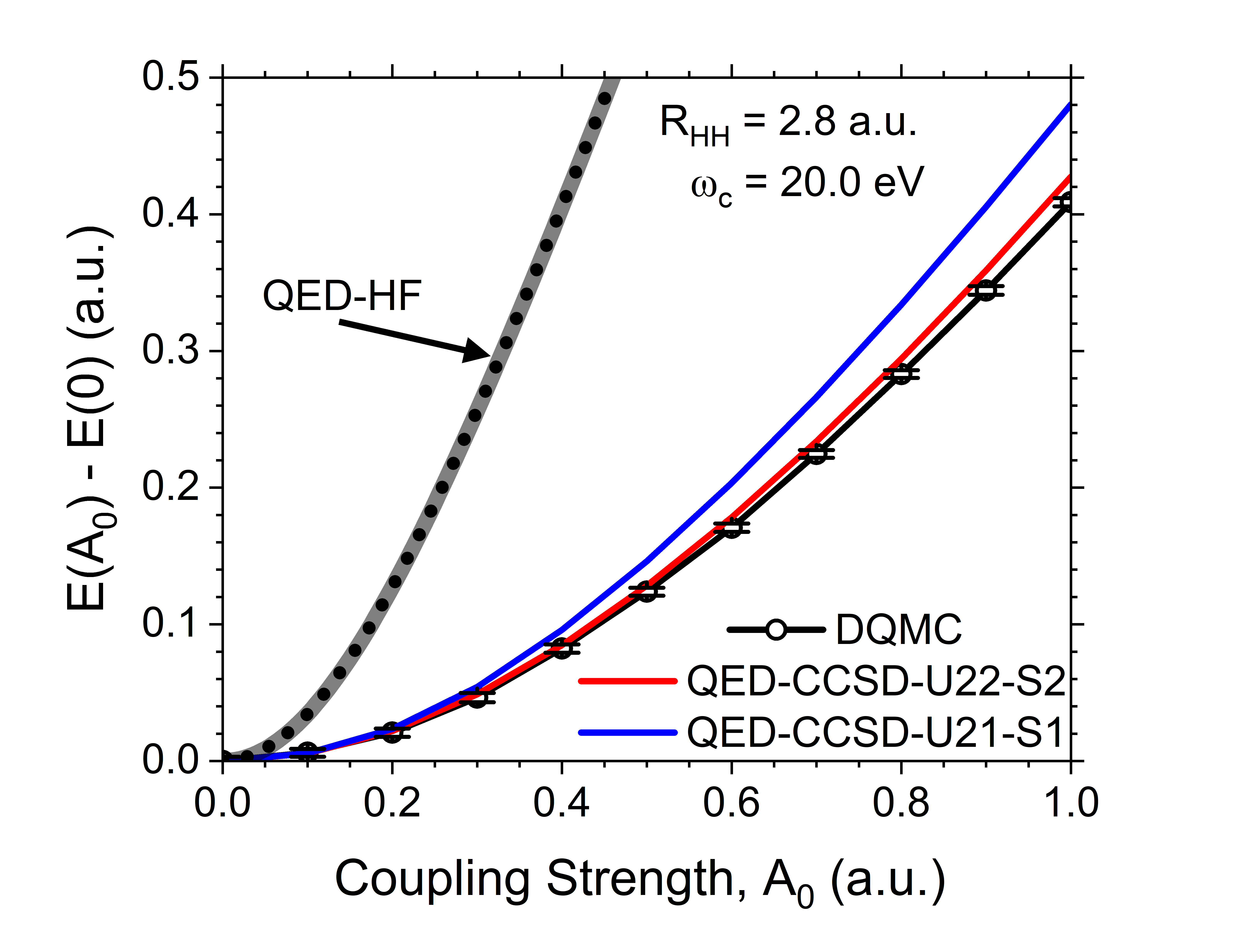}
  \caption{\footnotesize The ground state potential energy of the H$_2$ system as a function of light-matter coupling strength $A_0$. Four methods are compared: DQMC (current work, solid black curve and circles), QED-CCSD-U22-S2 (red curve), QED-CCSD-U21-S1 (blue curve), and QED-HF (dashed black, thick transparent black). The results of the QED-CCSD and QED-HF (thick transparent black) are computed using code published in Ref.~\citenum{Pavosevic2022JACS}, and the QED-HF (dashed black) is generated with an in-house code.\cite{Weight_PolaritonPCCP_2023Arxiv} For the DQMC method, since the plot presents the ground state energy with respect to the energy at zero coupling strength, the standard deviation of the difference of the means $\sigma_{A-B}$ are calculated as $\sigma_{A-B} = \sqrt{\sigma_{A}^2 + \sigma_{B}^2}$, where $\sigma_{A}$ and $\sigma_{B}$ are the standard deviations of the mean for $E(A_0)$ and $E(0)$, The cavity frequency and nuclear separation are fixed at $\omega_\mathrm{c}$ = 20.0 eV and $R_\mathrm{HH}$ = 2.8 a.u., respectively. 
  }
  \label{FIG:RUBIO_METHOD_COMP_Fixed_RHH}
\end{figure}

At this point, it is prudent to compare/benchmark the results of the DQMC for the polaritonic system with the state-of-the-art coupled cluster approach for polaritonic eigenstates, i.e., the self-consistent QED-CCSD method. This method encompasses diverse treatments of photonic excitation and coupled excitations. In this work, the notation QED-CCSD-U2n-Sm specifies the level of the coupled cluster approach, where $n$ represents the truncation level for the coupled photonic/electronic excitation and $m$ indicates pure photonic excitation. As outlined in Refs.~\citenum{White:2020jcp, Pavosevic2022JACS, alec2020jcp, Weight_PolaritonPCCP_2023Arxiv}, multiple strategies can be applied; we focus on those where $n = m$ and $ n \in \{1,2\} $. 

Figure~\ref{FIG:RUBIO_METHOD_COMP_Fixed_RHH}  presents the results of these two approaches along with the DQMC of the current work as well as the QED-HF result. The change in the  ground state energy, $E(A_0) - E(0)$, of the polaritonic system as a function of the light-matter coupling strength  $A_0$ is shown. The scQED-CCSD-U21-S1 approach includes only a single photonic excitation and its subsequent interaction with the electronic system. In contrast, the scQED-CCSD-U22-S2 approach includes double excitations of the cavity. This is expected to provide increased accuracy in a similar sense as the double electronic excitations increase the accuracy for the electronic ground state. Firstly, the QED-HF results significantly overestimate the change in ground state energy with increasing light-matter coupling strength. The DQMC approach agrees well with the scQED-CCSD-U22-S2 approach up to roughly $A_0 \sim 0.5$ a.u., beyond which the CCSD results exceed the DQMC's standard deviation. The scQED-CCSD-U21-S1 surpasses the DQMC's error bars around $ A_0 \approx 0.2 $ a.u. Given that the DQMC method captures exact correlations among electrons and between electrons and photons, we expect that a scQED-CCSD method with increased photonic excitation would align more closely with DQMC energies.

We next inspect the overlaps between the  photonic wavefunction and the Fock state basis in the cavity position representation, $c_f (A_0) = \langle \phi_\mathrm{ph}^\mathrm{DQMC}(A_0) | f \rangle$, where $f$ is the Fock state with $f$ photons and $|\phi_\mathrm{ph}^\mathrm{DQMC}(A_0)\rangle$ is the wavefunction as computed by the DQMC approach in this study (see Figs.~\ref{FIG:PHOTON_WFN_FOCK_BASIS},~\ref{FIG:PHOTON_WFN_FOCK_BASIS_EVEN_CONNECTED} in the \textbf{Supporting Information}). At  $A_0=$  0.0, 0.5, and 1.0 a.u., the amplitudes of the zero-photon Fock state are  $c_f$ =  1.0, 0.95, and 0.89, respectively.  At $A_0$ = 1.0~a.u., there are significant contributions from the higher photon-number Fock states, $c_f$ : 0.89 ( $f$ = 0 ), 0.37 ($f$ = 2), 0.20 ($f$ = 4), 0.12 ($f$ = 6), and 0.07 ($f$ = 8). It it worth noting that contributions from odd-photon-number Fock states are below 0.001, due to the ``even''-symmetry of the ground state wavefunction. Furthermore, over 99.5\% of the photonic wavefunction is encapsulated by the first 10 Fock states ($0 \le f \le 9$) for $A_0 = 1.0$ a.u. This quantitative analysis of the photonic wavefunction underscores the necessity for the scQED-CCSD approach to incorporate additional photonic excitations to capture the required correlations. This also elucidates the discrepancy between the DQMC and scQED-CCSD  results at large couplings.

It should be noted that utilizing unperturbed Fock/number states as the basis for photonic DOFs is a user choice. In principle, one may consider a polarized basis, such as the polarized Fock states\cite{Mandal2020JPCL} or the generalized coherent state basis.~\cite{philbin2014AmerJourPhys} These basis sets introduce a shift in the photonic coordinate potentially enabling a superior basis to the bare and unshifted Fock states presented in this work.  Additionally, by nature of the coupled cluster approach, higher excitation levels than the explicitly included excitation operators are always present due to the exponential anzatz. For example, the authors of Ref.~\citenum{Pavosevic2022JACS} utilized the shifted Fock basis\cite{haugland2020PRX} in the generation of the scQED-CCSD code used in this work. Thus the results shown in Fig.~\ref{FIG:RUBIO_METHOD_COMP_Fixed_RHH} should benefit from such transformation. Yet, some errors still exist due to the lack of sufficiently large number of photonic excitations.  

Here, we have shown that even the state-of-the-art coupled cluster approaches can be far from the exact solution even for simple systems  such as H$_2$. We suggest the DQMC scheme as a valuable tool for gaining insights into elusive correlations within various systems, particularly when testing novel polaritonic many-body techniques.

Returning to the potential energy surface, Figure~\ref{FIG:H2_DISS_Cavity_DIFFERENCE}  compares the potential energy surfaces in the cavity to the zero-coupling case, i.e, the differences represented as $E(A_0) - E(0)$. These quantities are shown across a range of light-matter coupling values for four distinct cavity frequencies: (a) 5.0, (b) 10.0, (c) 15.0, and (d) 20.0 eV. A noticeable maximum value emerges, especially in the case of large coupling. This maximum moves to shorter nuclear separation distances, $R_\mathrm{HH}$, with increasing light-matter coupling strength. This observation resembles the shift noticed in the minima of the potential energy surface, and it will become apparent in the observables we discuss later in this study. In fact, all of these alterations are closely connected to modifications in the electronic wavefunction, specifically the molecular quadrupole, which is the focus of the upcoming section.

\subsection{Polaritonic Wavefunctions}\label{SEC:WAVEFUNCTIONS}

\begin{figure}[t!]
\centering
\includegraphics[width=0.7\linewidth]{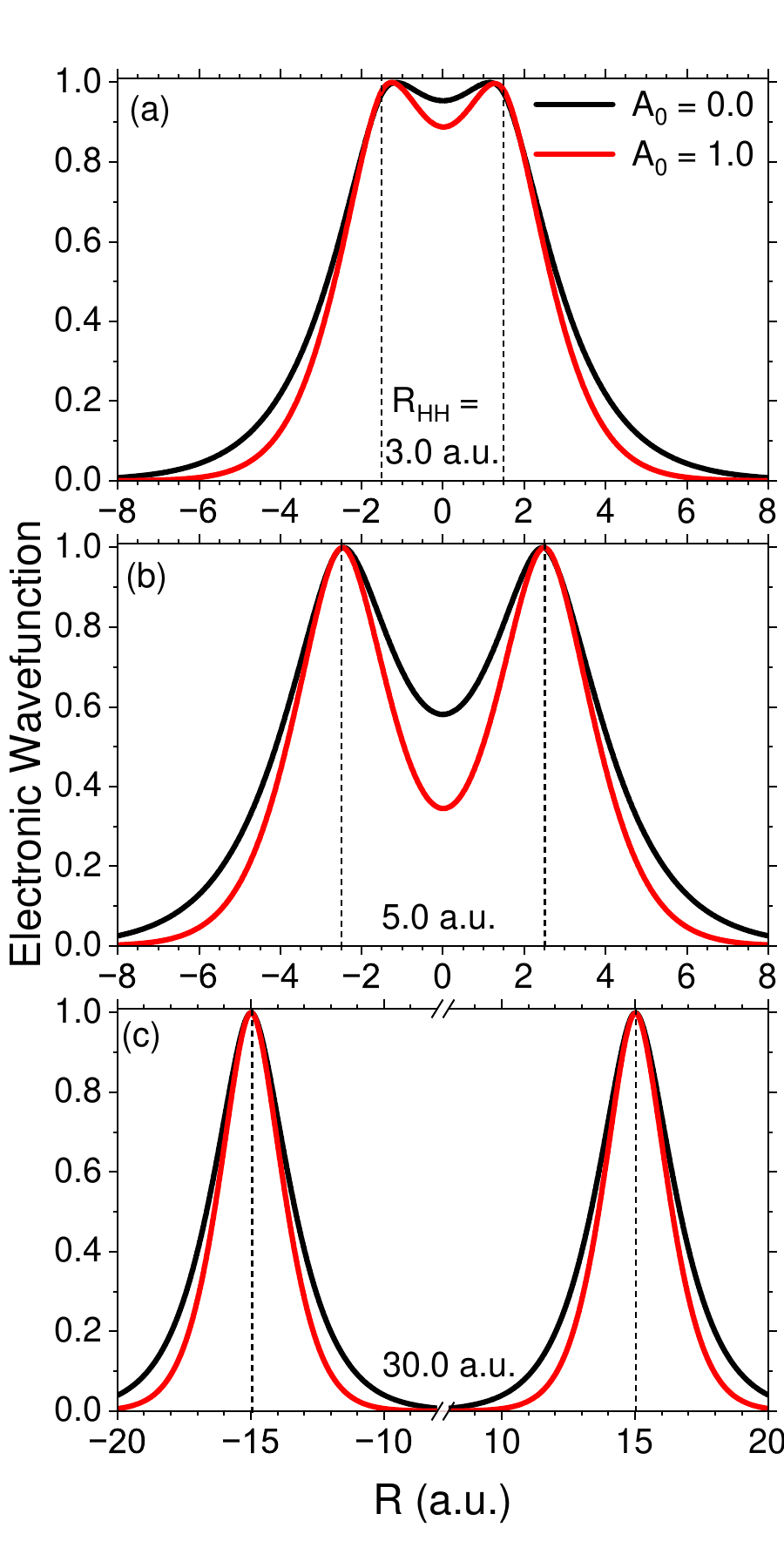}
  \caption{\footnotesize The electronic wavefunction for the H$_2$ system coupled to the cavity at various nuclear separation lengths $R_\mathrm{HH}$ = 3 (a), 5 (b), and 30 (c) a.u. for two coupling strengths $A_0$ = 0.0 (black) and 0.5 (red) a.u. The wavefunctions are all normalized by their maximum value for plotting purposes as $ \phi_\mathrm{el}^\mathrm{DQMC}/\mathrm{MAX}[\phi_\mathrm{el}^\mathrm{DQMC}]\rightarrow \phi_\mathrm{el}^\mathrm{DQMC}$. For all panels, the cavity frequency is set to $\omega_\mathrm{c}$ = 20.0 eV with cavity polarization being parallel to the bond axis.
  }
  \label{FIG:Electronic_Wavefunction}
\end{figure}

Frequently, there is an interest in properties pertaining to the electronic part of the wavefunction confined within the cavity. The alterations noted within the electronic subsystem provide direct information regarding chemical reactions. The photonic wavefunction's expansion in the Fock basis has been previously illustrated in Supplementary Figs.~\ref{FIG:PHOTON_WFN_FOCK_BASIS},~\ref{FIG:PHOTON_WFN_FOCK_BASIS_EVEN_CONNECTED}.  Here we focus on the electronic wavefunction $|\phi_\mathrm{el}^\mathrm{DQMC}\rangle$ plotted in Fig.~\ref{FIG:Electronic_Wavefunction} at various nuclear separation lengths  $R_\mathrm{HH}$. For each nuclear separation, the wavefunctions for two distinct coupling strengths, $A_0$ = 0.0 and 0.5 a.u. are shown.
A critical observation pertains to the influence of the cavity on the electronic wavefunction's localization. The underpinning for this behavior can be traced to the Pauli-Fierz Hamiltonian, which incorporates the molecular dipole operator via both direct light-matter interaction $\hat{H}_\mathrm{el-ph}$ and through the DSE $\hat{H}_\mathrm{DSE}$ term in Eq.~\ref{EQ:H_PF}. This implies that the wavefunction tends to become an eigenstate of the dipole (or position) operator, in the limit when $\hat{H}_\mathrm{el-ph},~\hat{H}_\mathrm{DSE} >> \hat{H}_\mathrm{el}$.  

To characterize the localization in quantum mechanical systems, the inverse participation ratio (IPR) is widely used as a quantitative metric.\cite{tretiak2002CR,weight2021JPCC,kilina_PolyFluorenes_ACSNano2008,weight_SWCNTRaman_JPCL2023,zheng_photoluminescence_ACSNano2021,wegner_IPR_ZPBCM1980,murphy_IPR_PRB2011} The IPR accesses the spread of a quantum mechanical wavefunction $\psi$ across its basis (in this case, real-space position $R$) and is defined as
\begin{equation}
    \mathrm{IPR} = \frac{1}{\sum_{j} P_j^2},~~~~~~P_j = \frac{|\psi(R_j)|^2}{\sum_{k} |\psi(R_k)|^2}
\end{equation}
Here, $|\psi(R_k)| = \psi(R_k) \equiv \phi_\mathrm{el}^\mathrm{DQMC}(R_k)$ in this work, since the wavefunction in the position representation is both real- and positive-valued. The resulting IPR spans a range between 1 and $N_\mathrm{basis}$, where $N_\mathrm{basis}$ is the number of basis states (or discrete positions $R_j$). When $\psi (R_j) = \delta_{R_j,R_0}$, IPR = 1, and when $\psi (R_j) = 1/N_\mathrm{basis}$, IPR = $N_\mathrm{basis}$. To convert this value into a spatial length, one can multiply by the grid spacing $\Delta R$ to obtain the localization length $L_d = \mathrm{IPR}\times \Delta R$.  

\begin{figure}[t!]
\centering
\includegraphics[width=1.0\linewidth]{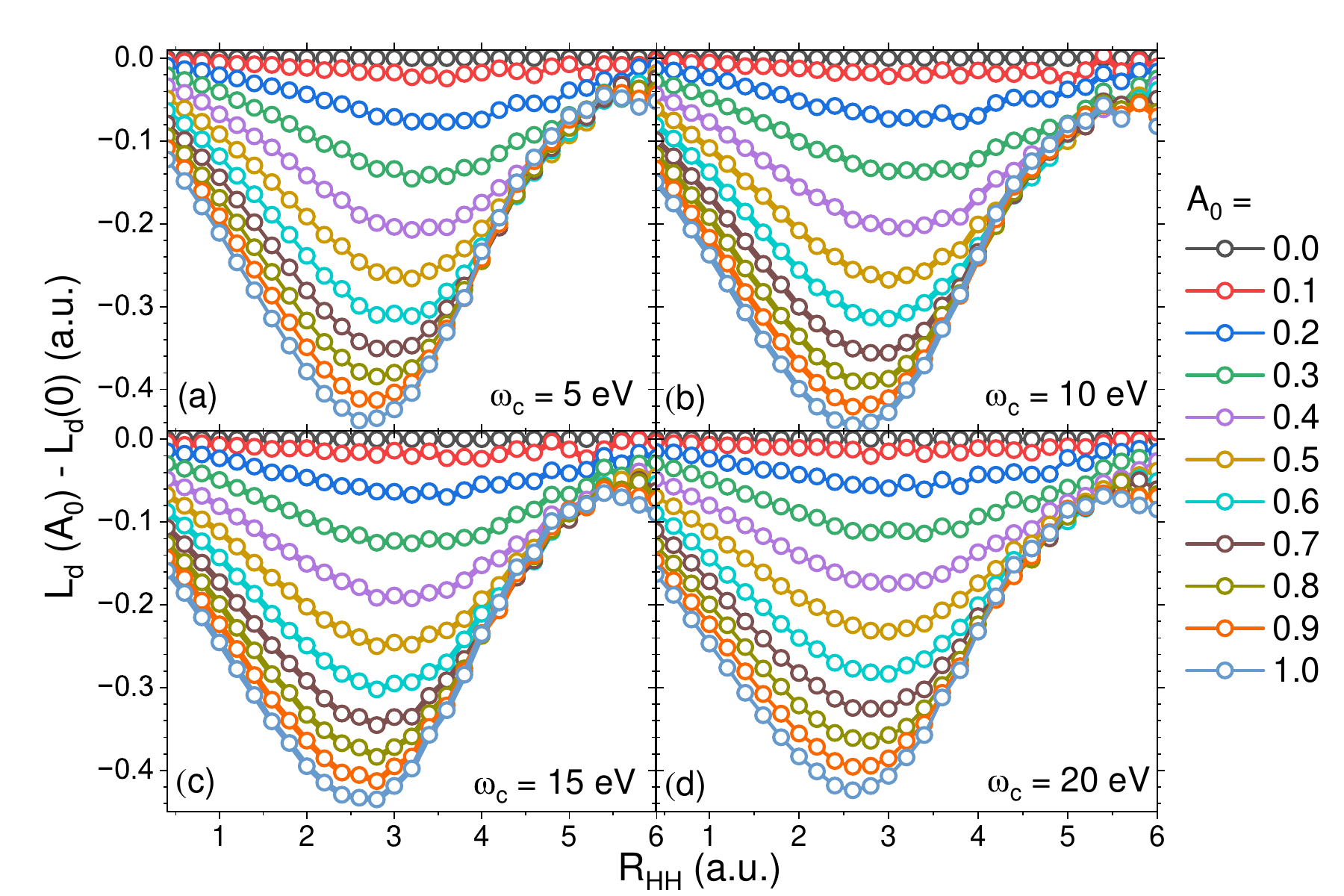}
  \caption{\footnotesize The difference in wavefunction localization of the ground state H$_2$  dissociation with respect to the pristine system (\textit{i.e.}, no cavity), $L_d(A_0) - L_d(0)$, at various coupling strengths $A_0$ (colors) for four cavity frequencies $\omega_\mathrm{c}$ = 5.0 (a), 10.0 (b), 15.0 (c), and 20.0 (d) eV.
  }
  \label{FIG:Ld_DIFF}
\end{figure}

Supplementary Figure~\ref{FIG:Ld} displays the computed localization length, $L_d$, for every DQMC simulation of the electronic wavefunction, projected along the bond axis. The most meaningful representation is rather the change in the localization with respect  to outside the cavity, $L_d(A_0) - L_d(0)$, which is presented in Fig. ~\ref{FIG:Ld_DIFF} for all light-matter coupling strengths $A_0$ and four cavity frequencies $\omega_\mathrm{c}$. There is a pronounced nuclear separation  $R_\mathrm{HH}$ where the localization of the wavefunction exhibits a maximal change compared to outside the cavity, which strongly depends on the light-matter coupling $A_0$. This trend mirrors our earlier observations regarding the potential energy surface minimum, as depicted in Fig.~\ref{FIG:PROPERTIES_3PANEL}a. The cavity frequency, again, has minimal effect on the localization of the electronic wavefunction. At a fixed nuclear separation of $R_\mathrm{HH}$ = 2.8 a.u. (near the maximum localization), the extent of enhanced wavefunction localization is presented in Fig.~\ref{FIG:PROPERTIES_3PANEL}b and follows a Gaussian function of the coupling strength. The data  shown in Fig.~\ref{FIG:PROPERTIES_3PANEL}b represent  an average over four cavity frequencies,  which provide increased statistical clarity. Our results are suggestive of a generic trend, wherein the electronic localization adheres to a Gaussian profile as a function of the coupling strength. Such Gaussian correlations with coupling strength have also been observed in relation to the splitting between ground and excited state avoided crossings for the LiF system\cite{Mandal2020JPCL}. These were attributed to a wavefunction overlap between two shifted Fock states  (\textit{i.e.}, a Gaussian involving the light-matter coupling strength) in the polarized Fock state representation.

\subsection{Average Photon Number}\label{SEC:PHOTONIC_OCCUPATION}

\begin{figure}[t!]
\centering
\includegraphics[width=1.1\linewidth]{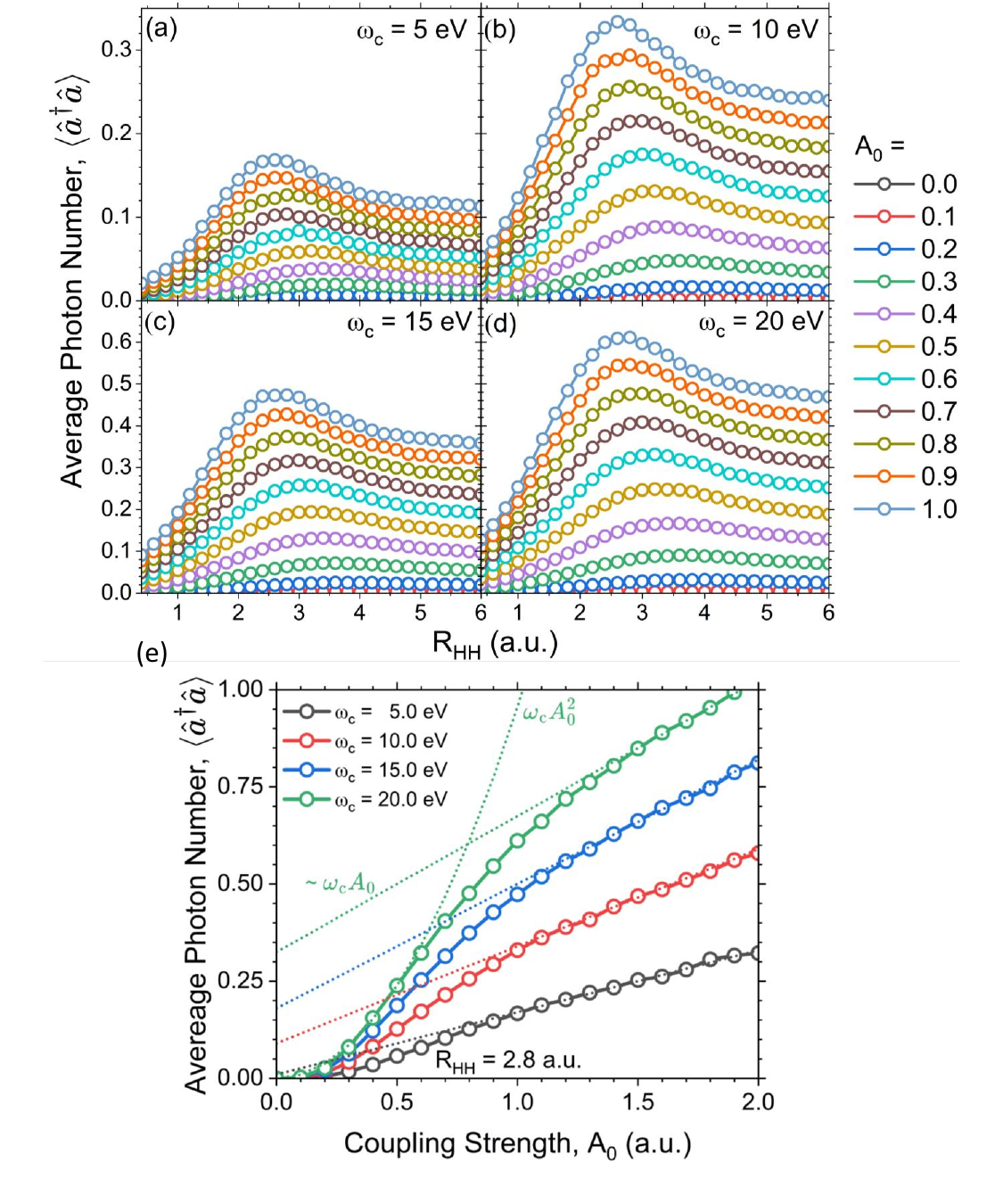}
  \caption{\footnotesize The average photon number, $\langle \hat{a}^\dag\hat{a} \rangle$, throughout the ground state H$_2$ dissociation with respect to the uncoupled system, $L_d(A_0) - L_d(0)$, at various coupling strengths $A_0$ (colors) for four cavity frequencies $\omega_\mathrm{c}$ = 5.0 (a), 10.0 (b), 15.0 (c), and 20.0 (d) eV. (e) The average photon number, $\langle \hat{a}^\dag\hat{a} \rangle$, as a function of light-matter coupling strength $A_0$ for a fixed nuclear separation $R_\mathrm{HH}$ = 2.8 a.u. for four cavity frequencies $\omega_\mathrm{c}$ = 5.0 (black), 10.0 (red), 15.0 (blue), and 20.0 (green) eV. The dotted curves present the scaling at the low (\textit{i.e.}, quadratic) and high (\textit{i.e.}, linear) coupling regimes.
  }
  \label{FIG:PHOTON_NUMBER}
\end{figure}

The final observable of interest is the average photon number in the ground state. This is a direct extension of our earlier discussion about the overlap of the DQMC photonic wavefunction with the number (or Fock) basis (see Supplementary Fig.~\ref{FIG:PHOTON_WFN_FOCK_BASIS_EVEN_CONNECTED}). In the photon number basis, the expression $\langle \hat{a}^\dag\hat{a} \rangle$ represents the average occupation of the mode in the Pauli-Fierz (or length) gauge, which may differ from that in the Coulomb (or ``$p\cdot A$'') gauge. Nevertheless, the average occupation can undergo a unitary rotation to any other gauge,\cite{mandal_QEDChemRev_Arxiv2022,Hu_ML_JCTC2023} and we expect that the absolute magnitudes of the photon number might differ, but the physical trends should remain unchanged.

Figure~\ref{FIG:PHOTON_NUMBER}a-d presents the average photon number $\langle \hat{a}^\dag\hat{a} \rangle$ throughout the H$_2$  dissociation, for various  light-matter coupling strengths $A_0$ and for four cavity frequencies $\omega_\mathrm{c}$. Analogous to the previous observables, such as the change in electronic wavefunction localization $L_d$, there is a distinct peak in photon number at a certain value of $R_\mathrm{HH}$, depending on the coupling strength $A_0$. The peak's location is illustrated in Fig.~\ref{FIG:PROPERTIES_3PANEL}c and fit to a linear function, where the data has been averaged over the four cavity frequencies $\omega_\mathrm{c}$.

At a fixed $R_\mathrm{HH}$ = 2.8 a.u., as depicted in Fig.~\ref{FIG:PHOTON_NUMBER}e, the average photon number behaves intriguingly. For smaller light-matter coupling strengths $A_0 \leq 0.5$ a.u., the photon number increases quadratically, $\langle \hat{a}^\dag\hat{a}\rangle \sim \omega_\mathrm{c} A_0^2$, with its leading coefficient dependent on the cavity frequency. Conversely, for larger light-matter coupling $A_0 \geq 0.5$ a.u., the average photon number increases linearly (up to the maximal coupling strength computed in this work) with slopes dependent of the cavity frequency, $\langle \hat{a}^\dag\hat{a}\rangle \sim \omega_\mathrm{c} A_0$. The convergence of the average photon number in the ground state is tested with respect to simulation timestep, as shown in Supplementary Fig.~\ref{FIG:Photon_Number_Convergence_Timestep}. Given these substantial changes in the photon number, there exists potential for fascinating quantum measurements in highly entangled many-body states, especially in dynamic scenarios where complex interplay between the  nuclear and photonic DOFs can induce significant fluctuations in the photon number, even in the ground polaritonic state.

\section{Conclusions}
In  summary, this work presents a diffusion quantum Monte Carlo (DQMC)  scheme for direct simulations of molecular polaritons, which necessarily includes complicated correlations between the electronic, nuclear, and photonic degrees of freedom. This scheme represents a thoroughly \textit{ab initio} approach, which is formally exact for the two-electron system interacting with, in principle, infinite cavity modes. As a testament to its capabilities, we study the H$_2$ molecular dissociation coupled to a single cavity mode. The results are compared directly with state-of-the-art quantum electrodynamical (QED) coupled cluster  approaches, which is a formally exact solution to the pristine molecular system (in the absence of cavity). We emphasize that the accuracy of the latter method is largely contingent upon the truncation of photonic excitation in the exponential ansatz. Our comparative analysis evidence that even the highest fidelity approach currently available fails in even the simplest molecular system when the light-matter coupling  strengths become large enough to necessarily include contributions from photonic excitations beyond one or two photonic number (or Fock) states.

The DQMC approach is then used to explore various observables for the H$_2$ system, such as the localization extent of the electronic and the decomposition of the photonic wavefunctions in the Fock basis. Additionally, linear trends in the light-matter coupling strength are identified for the nuclear separations which exhibit (I) the minimum of the potential energy surfaces and (II) the maximum average photon number. Moreover, a Gaussian trend emerges in the changes related to wavefunction localization. In the former two cases, the change of the properties with respect to the nuclear separation length are directly relevant to modifications of chemical reactions, where even the smallest nuclear displacement may alter reaction pathways and resulting products. Meanwhile, the wavefunction's localization can offer insights into processes demanding wavefunction overlap, such as Dexter energy transfer or H/J-aggregates (interacting excitonic systems). This holds implications for exciton-polariton transport in molecular and solid-state materials.

Altogether DQMC  represents a promising route toward the direct and accurate simulation of simple systems, where the results are exact. Applications to more intricate systems are possible as well by leveraging the fixed node approximation for many-electron systems and excited states.\cite{foulkes_quantum_2001,kulahlioglu_multistate_2023,umrigar_alleviation_2007,kim_qmcpack_2018}  We note that in the realm of polaritonic DQMC, the fixed node approximation needs to be carefully implemented, since the nodal surface will likely undergo changes as a result of interaction with the cavity . Thus, using a mean-field wavefunction (\textit{i.e.}, Hartree-Fock) must either (I)  include the cavity effects on the single-particle orbitals  (\textit{i.e.}, QED-HF)\cite{liebenthal_assessing_2023} or (II) allow the nodal surface  to relax in order to minimize the energy while retaining the correct electronic statistics. These ideas direct our future studies in this area.

Presented DQMC approach can be further readily employed to explore polaron formation and its inherent properties, including polaron radius and binding energy, given the bosonic nature of both photons and phonons. Most importantly, and without loss of accuracy, the many-molecule and many-cavity-mode systems can be directly simulated, assuming (I)  the molecules do not interact directly via Coulomb potential (\textit{i.e.}, only through mutual interaction with the cavity) and (II) the cavity modes do not include the complex-valued nature of the Pauli-Fierz Hamiltonian in the absence of the long-wavelength approximation. Despite these constraints, our approach illuminates solutions for simple molecular systems inside the cavity. Further, the extension toward plasmonic cavities, where the  light-matter coupling strength depends on position, $A_0 \rightarrow A_0({\bf r})$,  would be straightforward with minimal changes to the code. DQMC's relative simplicity and favorable scaling with dimensionality, typically a constraint for advanced methods like coupled cluster, cements its place as a promising tool for investigating \textit{ab initio} molecular systems inside cavities.

\section*{Author Contributions}
We strongly encourage authors to include author contributions and recommend using \href{https://casrai.org/credit/}{CRediT} for standardised contribution descriptions. Please refer to our general \href{https://www.rsc.org/journals-books-databases/journal-authors-reviewers/author-responsibilities/}{author guidelines} for more information about authorship.

\section*{Conflicts of interest}
There are no conflicts to declare.

\section*{Acknowledgements}
The authors acknowledge the support from the Laboratory Directed Research and Development Funds (LDRD) at Los Alamos National Laboratory (LANL) and the US DOE, Office of Science, Basic Energy Sciences, Chemical Sciences, Geosciences, and Biosciences Division under Triad National Security, LLC (``Triad") contract Grant 89233218CNA000001 (FWP: LANLECF7).  The research is performed in part at the Center for Integrated Nanotechnologies (CINT), a U.S. Department of Energy, Office of Science user facility at LANL. LANL is operated by Triad National Security, LLC, for the National Nuclear Security Administration of the U.S. Department of Energy (Contract No. 89233218CNA000001). Computing resources were provided by the Center for Integrated Research Computing (CIRC) at the University of Rochester. 





\bibliography{main} 
\bibliographystyle{rsc} 

\clearpage

\onecolumn

{\centering \bf \Huge  Supporting Information}

\setcounter{section}{0}
\setcounter{subsection}{0}
\setcounter{subsubsection}{0}
\setcounter{equation}{0}
\setcounter{figure}{0}

\renewcommand{\theequation}{S\arabic{equation}}
\renewcommand{\thesection}{S\arabic{section}}
\renewcommand\thefigure{S\arabic{figure}}

\clearpage

\begin{figure}
\centering
\includegraphics[width=0.8\linewidth]{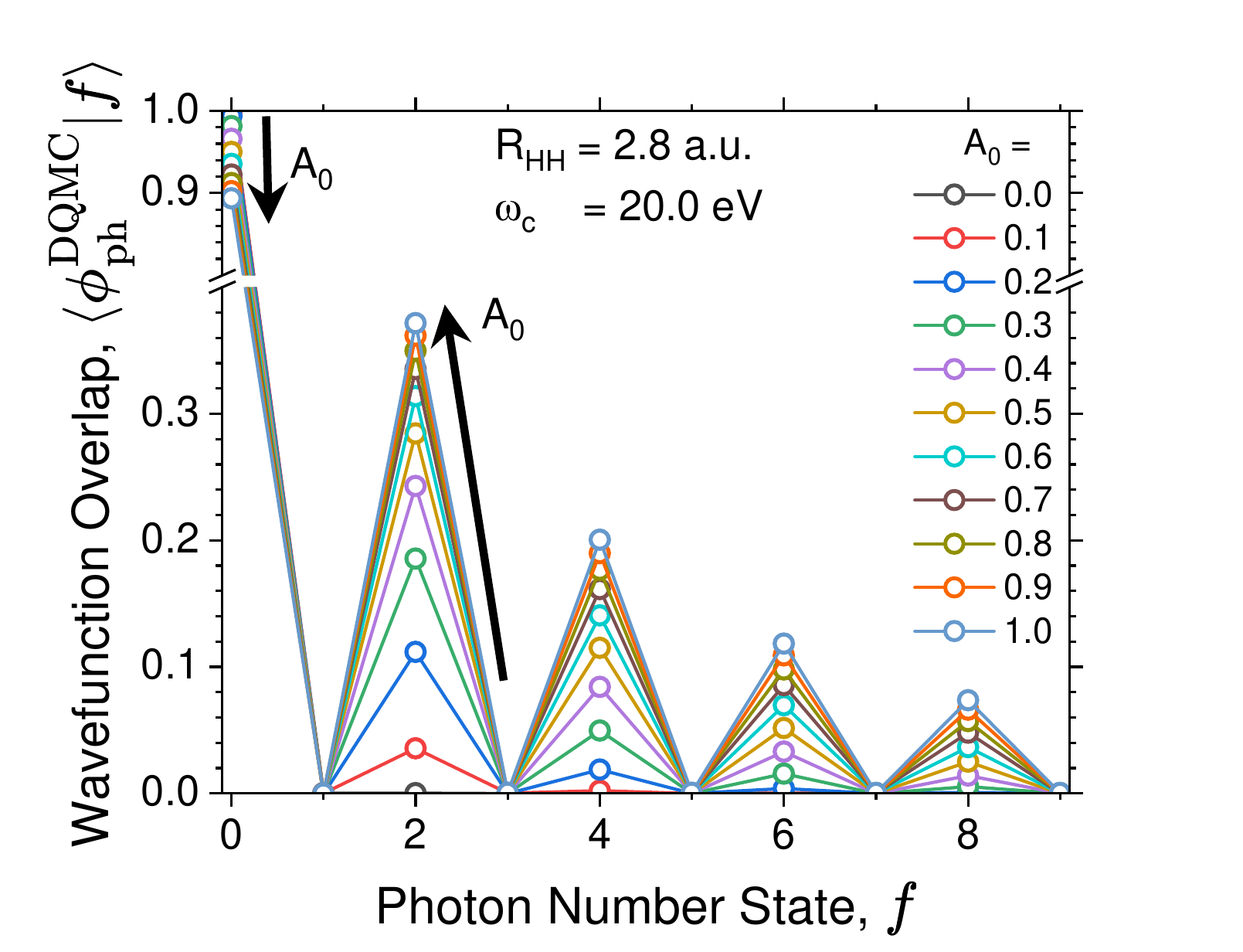}
  \caption{Photonic number state (or Fock state) expansion of the photonic DQMC wavefunction at various coupling strengths $A_0 \in$ (0.0, 1.0) a.u. for fixed cavity frequency $\omega_\mathrm{c}$ = 20.0 eV and nuclear separation length $R_\mathrm{HH}$ = 2.8 a.u.
  }
  \label{FIG:PHOTON_WFN_FOCK_BASIS}
\end{figure}

\clearpage

\begin{figure}
\centering
\includegraphics[width=0.8\linewidth]{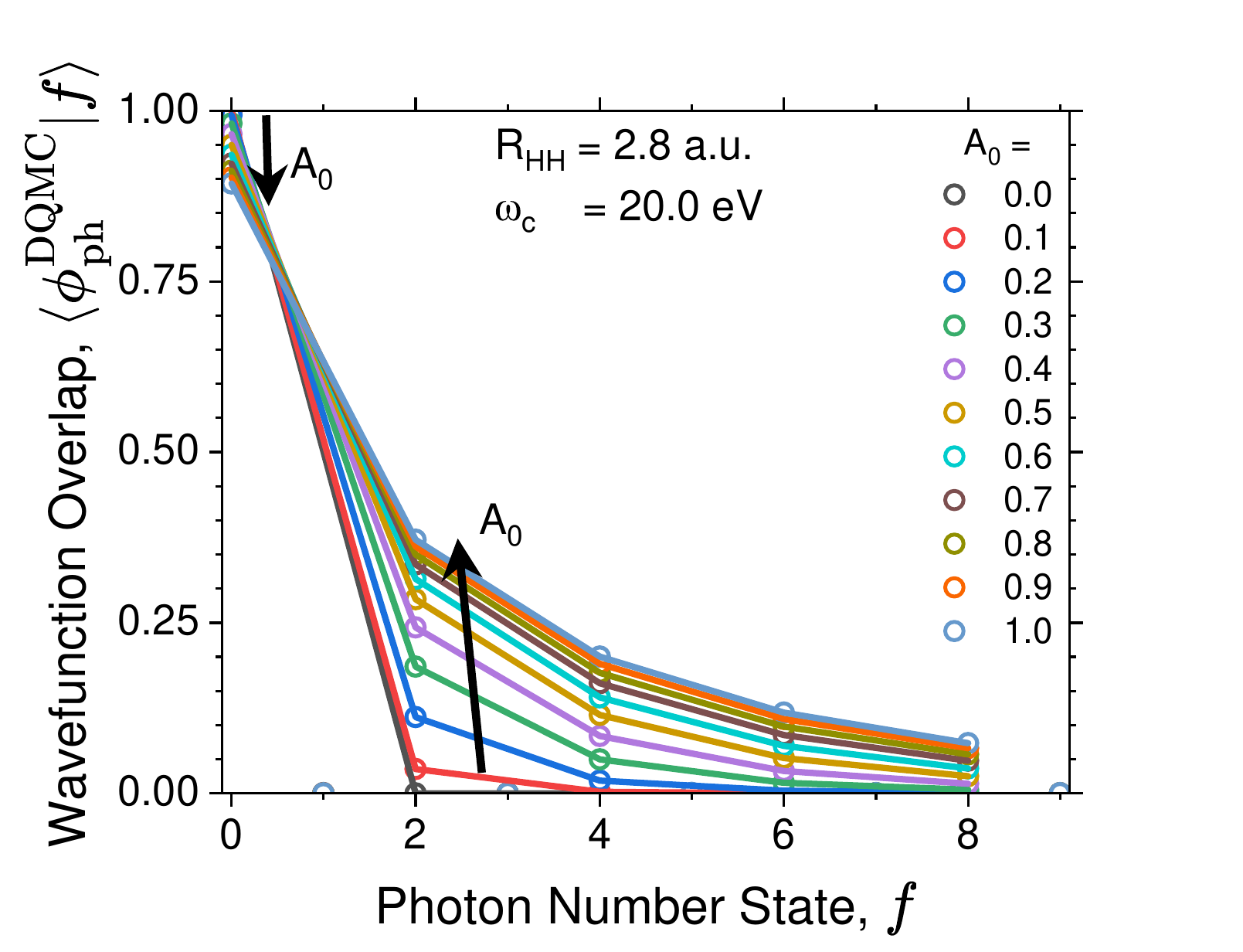}
  \caption{Photonic number state (or Fock state) expansion of the photonic DQMC wavefunction at various coupling strengths $A_0 \in$ (0.0, 1.0) a.u. for fixed cavity frequency $\omega_\mathrm{c}$ = 20.0 eV and nuclear separation length $R_\mathrm{HH}$ = 2.8 a.u.
  }
  \label{FIG:PHOTON_WFN_FOCK_BASIS_EVEN_CONNECTED}
\end{figure}

\clearpage

\begin{figure*}[t!]
\centering
\includegraphics[width=0.9\linewidth]{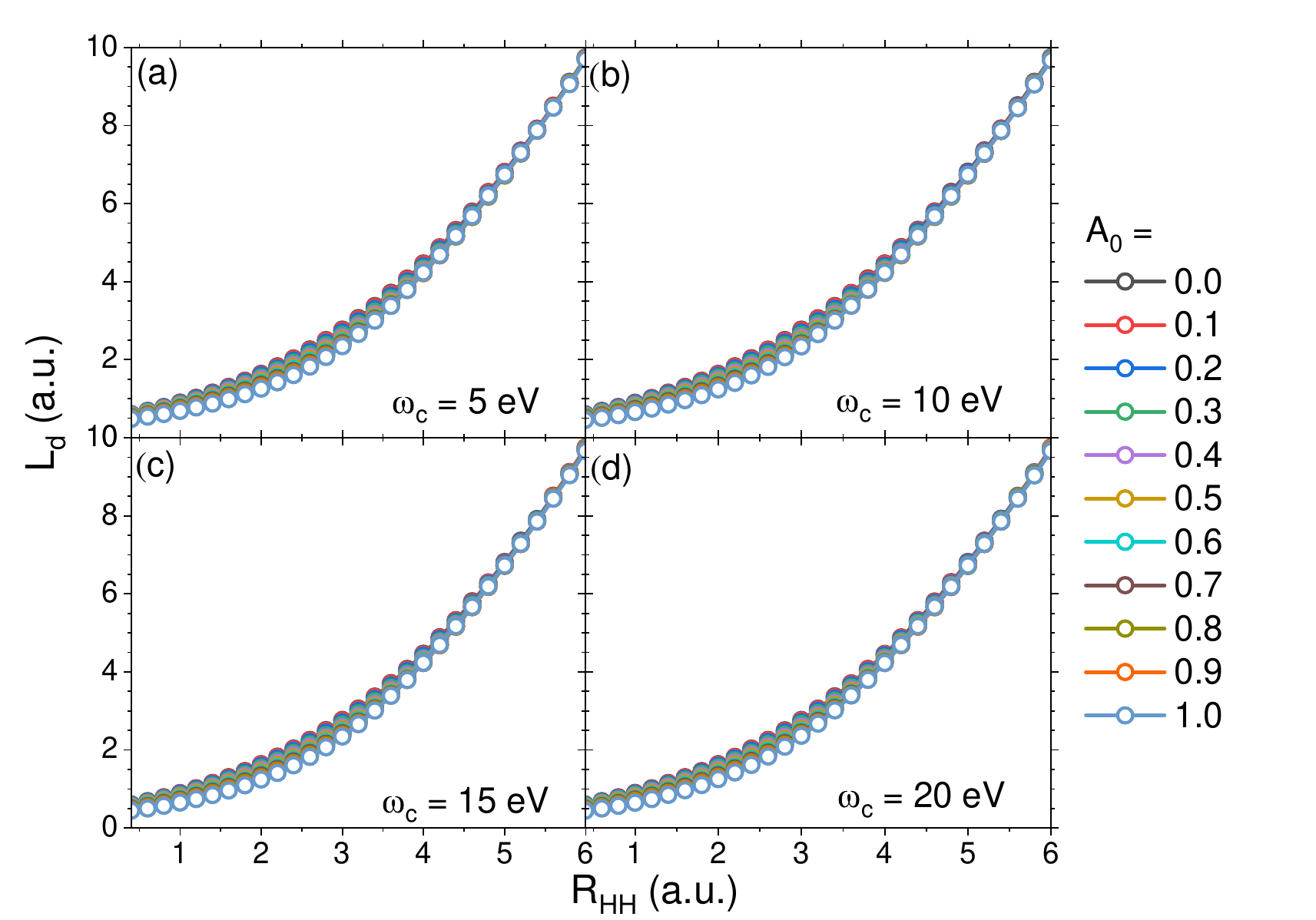}
  \caption{The wavefunction localization of the ground state H$_2$ dissociation, $L_d(A_0)$, at various coupling strengths $A_0$ (colors) for four cavity frequencies $\omega_\mathrm{c}$ = (a) 5.0, (b) 10.0, (c) 15.0, and (d) 20.0 eV.
  }
  \label{FIG:Ld}
\end{figure*}

\begin{figure}[t!]
\centering
\includegraphics[width=0.85\linewidth]{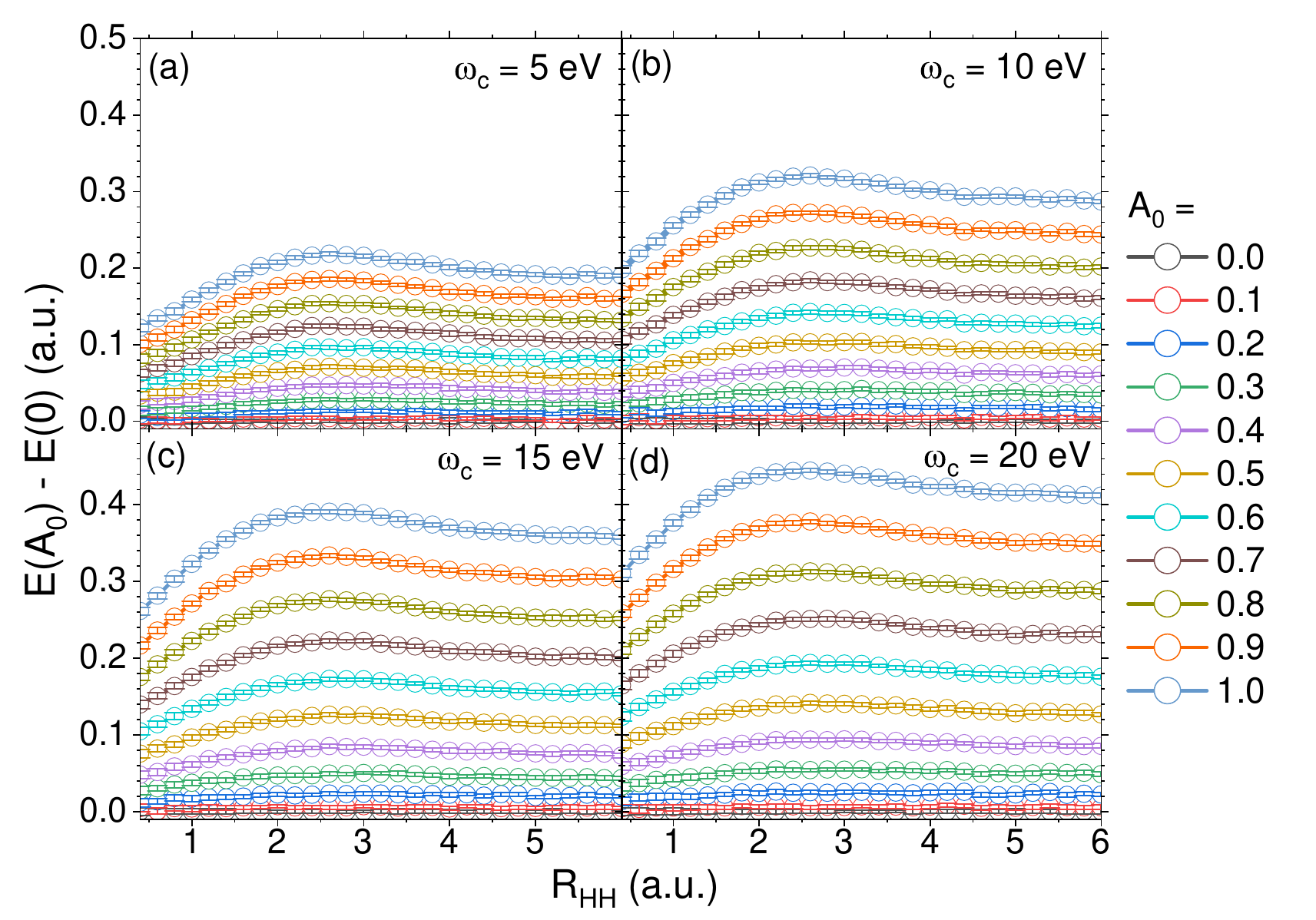}
  \caption{The energy difference of the ground state H$_2$ dissociation with respect to the uncoupled system (\textit{i.e.}, $A_0$ = 0.0 a.u.) at various coupling strengths $A_0$ = 0.0, 0.1, 0.2, 0.3, 0.4, and 0.5 a.u. for four cavity frequencies $\omega_\mathrm{c}$ = 5.0 (a), 10.0 (b), 15.0 (c), and 20.0 (d) eV. The cavity polarization is parallel to the bond axis. The standard deviation of the difference of the means $\sigma_{A-B}$ were calculated as $\sigma_{A-B} = \sqrt{\sigma_{A}^2 + \sigma_{B}^2}$, where $\sigma_{A}$ and $\sigma_{B}$ are the standard deviations of the mean for $E(A_0)$ and $E(0)$ at a fixed nuclear configuration.
  }
  \label{FIG:H2_DISS_Cavity_DIFFERENCE}
\end{figure}

\begin{figure*}[t!]
\centering
\includegraphics[width=0.5\linewidth]{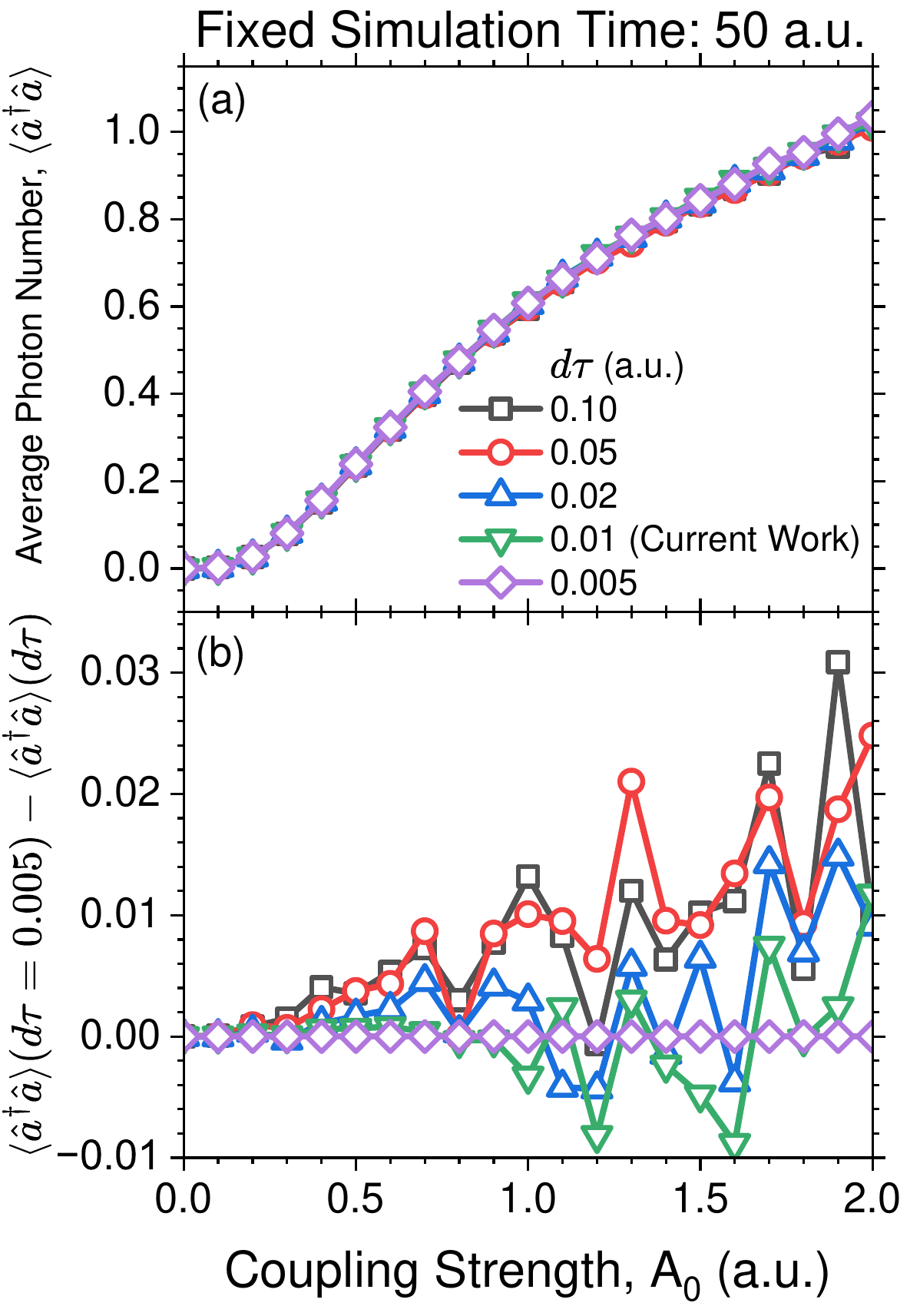}
  \caption{\footnotesize Convergence of the average photon number, $\langle \hat{a}^\dag\hat{a} \rangle$ as a function of light-matter coupling strength $A_0$ for a fixed nuclear separation $R_\mathrm{HH}$ = 2.8 a.u. for a variety of simulation timesteps: $d\tau$ = 0.10 (black squares), 0.05 (red circle), 0.02 (blue up triangles), 0.01 (green down triangles), and 0.005 a.u. (purple diamonds). Here, the cavity frequency is $\omega_\mathrm{c} = 20.0$ eV, and the simulation time was kept fixed at 50.0 a.u. All data in the main text was performed with d$\tau$ = 0.01 a.u.. The data presented here for d$\tau$ = 0.01 a.u. is repeated from from Fig.~\ref{FIG:PHOTON_NUMBER}e in the main text and exhibits less than 1\% deviation from d$\tau$ = 0.005 a.u. and less than 3\% deviation increasing the timestep to d$\tau$ = 0.10 a.u.
  }
  \label{FIG:Photon_Number_Convergence_Timestep}
\end{figure*}

\end{document}